\documentclass[zpreprint,zbstdefault,british]{./inputs/zeus_Kpaper}
\usepackage{./inputs/zeusdet_units}
\usepackage{pdflscape}
\chardef\usc=95
\chardef\til=126
\catcode`\@=11 % @ signs are now treated as letters
\DeclareRobustCommand\xdotspace{\futurelet\@let@token\@xdotspace}
\def\@xdotspace{%
  \ifx\@let@token.\else
  \ifx\@let@token\bgroup.\else
  \ifx\@let@token\egroup.\else
  \ifx\@let@token\/.\else
  \ifx\@let@token\ .\else
  \ifx\@let@token~.\else
  \ifx\@let@token!.\else
  \ifx\@let@token,.\else
  \ifx\@let@token:.\else
  \ifx\@let@token;.\else
  \ifx\@let@token?.\else
  \ifx\@let@token/.\else
  \ifx\@let@token'.\else
  \ifx\@let@token).\else
  \ifx\@let@token-.\else
  \ifx\@let@token\@xobeysp.\else
  \ifx\@let@token\space.\else
  \ifx\@let@token\@sptoken.\else
   .\space
   \fi\fi\fi\fi\fi\fi\fi\fi\fi\fi\fi\fi\fi\fi\fi\fi\fi\fi}
\catcode`\@=12 % @ signs are no longer letters
\newcommand{\CL}[1]{$#1\%$~C.L\xdotspace}
\newcommand{\stru}[2]{%
   \relax\ifmmode\hbox{\vrule height#1 depth#2 width0pt}%
   \else\vrule height#1 depth#2 width0pt\fi}
\newcommand{\Ronum}[1]{\uppercase\expandafter{\romannumeral#1}}
\newcommand{\ronum}[1]{\expandafter{\romannumeral#1}}
\DeclareRobustCommand{\LaTeXZ}{%
  \LaTeX\kern-.05em4\kern-.1em
  {\raisebox{-0.2ex}{$\scriptstyle\text{ZEUS}$}}\xspace}
\newcommand{\fig}[1]{Fig.~\ref{fig-#1}}

\newcommand{\tab}[1]{Table~\ref{tab-#1}}

\newcommand{\slashfrac}[2]{%
  \raisebox{0.5ex}{\ensuremath #1}\kern-0.12em/\kern-0.08em
  \raisebox{-.8ex}{\ensuremath #2}}
\newcommand{\sqr}[3]{%
    {\vcenter{\hrule height.#3ex\hbox{\vrule width.#2ex height#1ex
     \kern#1ex\vrule width.#3ex}\hrule height.#2ex}}}

\catcode`\@=11 % @ signs are now treated as letters
\newcommand{\parenbar}{\mathpalette\p@renb@r}
\def\p@renb@r#1#2{\vbox{%
  \ifx#1\scriptscriptstyle \dimen@.7em\dimen@ii.2em\else
  \ifx#1\scriptstyle \dimen@.8em\dimen@ii.25em\else
  \dimen@1em\dimen@ii.4em\fi\fi \offinterlineskip
  \ialign{\hfill##\hfill\cr
    \vbox{\hrule width\dimen@ii}\cr
    \noalign{\vskip-.3ex}%
    \hbox to\dimen@{$\mathchar300\hfil\mathchar301$}\cr
    \noalign{\vskip-.3ex}%
    $#1#2$\cr}}}
\catcode`\@=12 % @ signs are no longer letters

\ifzeusunit
  
\else
  
\fi

\newcommand{\IP}{{\rm I$\kern-0.01667em$P}\xspace}

\mathchardef\qsm=63
\mathchardef\pls=43
\mathchardef\mns=512
\mathchardef\plm=518
\mathchardef\eql=61
\mathchardef\smallleft=300
\mathchardef\smallright=301
\mathchardef\les=316
\mathchardef\gre=318
\mathchardef\leq=532
\mathchardef\grq=533
\catcode`\@=11 % @ signs are now treated as letters
\newcounter{pict@width}
\newcounter{pict@height}
\newlength{\pict@scale}
\newcommand{\psfigadd}[4]{%
\setcounter{pict@width}{1*\ratio{#2+\pict@scale/2}{\pict@scale}}
\setcounter{pict@height}{1*\ratio{#3+\pict@scale/2}{\pict@scale}}
\setlength{\unitlength}{\pict@scale}
\hbox to #2{\hspace{-\fill}\begin{picture}(\thepict@width,\thepict@height)
\put(0,0){\psfig{figure=#1,width=#2,height=#3,clip=}}
\SetScale{0.283466457}
\SetWidth{1.763889}
{#4}
\end{picture}}
}
\newcounter{pict@widthfst}
\newcounter{pict@widthscd}
\newcounter{pict@widthtot}
\newcommand{\psfigaddtwo}[7]{%
\setcounter{pict@widthfst}{1*\ratio{#2+\pict@scale/2}{\pict@scale}}
\setcounter{pict@widthscd}{1*\ratio{#2+#4+\pict@scale/2}{\pict@scale}}
\setcounter{pict@widthtot}{1*\ratio{#2+#4+#6+\pict@scale/2}{\pict@scale}}
\setcounter{pict@height}{1*\ratio{#3+\pict@scale/2}{\pict@scale}}
\setlength{\unitlength}{\pict@scale}
\hbox{\hspace{-\fill}\begin{picture}(\thepict@widthtot,\thepict@height)
\put(0,0){\psfig{figure=#1,width=#2,height=#3,clip=}}
\put(\thepict@widthscd,0){\psfig{figure=#5,width=#6,height=#3,clip=}}
\SetScale{0.283466457}
\SetWidth{1.763889}
{#7}
\end{picture}}
}
\newcommand{\psfigror}[4]{%
\setcounter{pict@width}{1*\ratio{#2+\pict@scale/2}{\pict@scale}}
\setcounter{pict@height}{1*\ratio{#3+\pict@scale/2}{\pict@scale}}
\setlength{\unitlength}{\pict@scale}
\hbox{\begin{picture}(\thepict@width,\thepict@height)
\put(0,\thepict@height){\psfig{figure=#1,width=#3,height=#2,clip=,angle=270}}
\SetScale{0.283466457}
\SetWidth{1.763889}
{#4}
\end{picture}}
}
\newcommand{\psfigrol}[4]{%
\setcounter{pict@width}{1*\ratio{#2+\pict@scale/2}{\pict@scale}}
\setcounter{pict@height}{1*\ratio{#3+\pict@scale/2}{\pict@scale}}
\setlength{\unitlength}{\pict@scale}
\hbox{\begin{picture}(\thepict@width,\thepict@height)
\put(0,0){\psfig{figure=#1,width=#3,height=#2,clip=,angle=90}}
\SetScale{0.283466457}
\SetWidth{1.763889}
{#4}
\end{picture}}
}
\catcode`\@=12 % @ signs are no longer letters
\newlength\listtextwidth

\catcode`\@=11 % @ signs are now treated as letters
\newlength{\@tabfninsert}
\newlength{\@tabfnwidth}
\newcommand{\tabfootnote}[2]{%
  \setlength{\@tabfninsert}{0.8em}
  \setlength{\@tabfnwidth}{\textwidth}
  \addtolength{\@tabfnwidth}{-\@tabfninsert}
  \addtolength{\@tabfnwidth}{-0.4em}
  \noindent\makebox[\@tabfninsert][r]{\footnotesize$^{#1}$\hfil}\hfill%
  \parbox[t]{\@tabfnwidth}{\footnotesize #2\hfill}}
\catcode`\@=12 % @ signs are no longer letters

\begin{document}
%------------------------------------------------------------------------------
%       Title sheet
%------------------------------------------------------------------------------
%
\prepnum{DESY 19-022}
% \prepdate{February 2019}          
% \prepnum{}
\prepdate{}          

\zeustitle{%
Limits on contact interactions\\
and leptoquarks at HERA
}

\zeusauthor{ZEUS Collaboration}

% Set \zeusdate to empty for final preprint.
\zeusdate{}

\maketitle

\begin{abstract}\noindent

High-precision HERA data corresponding to a luminosity of around
1\,fb$^{-1}$ have been used in the framework of $eeqq$ contact
interactions (CI) to set limits on possible high-energy contributions
beyond the Standard Model to electron--quark scattering. Measurements
of the inclusive deep inelastic cross sections in neutral and charged
current $ep$ scattering were considered. The analysis of the
$ep$ data has been based on simultaneous fits of parton distribution
functions including contributions of CI couplings to $ep$
scattering. Several general CI models and scenarios with heavy
leptoquarks were considered. Improvements in the description of the
inclusive HERA data were obtained for a few models.
Since a statistically significant deviation from the Standard Model
cannot be established, limits in the TeV range were set on all
models considered.

\end{abstract}

\thispagestyle{empty}
\cleardoublepage
%
%------------------------------------------------------------------------------
%       Author list
%------------------------------------------------------------------------------
%
\pagenumbering{roman}
\begin{center}
{                      \Large  The ZEUS Collaboration              }
\end{center}

{\small\raggedright

%  members:

H.~Abramowicz$^{26, r}$, 
I.~Abt$^{21}$, 
L.~Adamczyk$^{7}$, 
M.~Adamus$^{33}$, 
R. Aggarwal$^{3, b}$, 
S.~Antonelli$^{1}$, 
V.~Aushev$^{18}$, 
O.~Behnke$^{9}$, 
U.~Behrens$^{9}$, 
A.~Bertolin$^{23}$
% ...
\newpage

%       institutes:

{\setlength{\parskip}{0.4em}
\makebox[3ex]{$^{1}$}
\begin{minipage}[t]{14cm}
\textit{ INFN Bologna, Bologna, Italy}~$^{A}$

\end{minipage}

\makebox[3ex]{$^{2}$}
\begin{minipage}[t]{14cm}
\textit{ Physikalisches Institut der Universit\"at Bonn,
Bonn, Germany}~$^{B}$

\end{minipage}

\makebox[3ex]{$^{3}$}
\begin{minipage}[t]{14cm}
\textit{ Panjab University, Department of Physics, Chandigarh, India}

\end{minipage}

\makebox[3ex]{$^{4}$}
\begin{minipage}[t]{14cm}
\textit{ Calabria University,
Physics Department and INFN, Cosenza, Italy}~$^{A}$

\end{minipage}

\makebox[3ex]{$^{5}$}
\begin{minipage}[t]{14cm}
\textit{ National Centre for Particle Physics, Universiti Malaya, 50603 Kuala Lumpur, Malaysia}~$^{C}$

\end{minipage}

\makebox[3ex]{$^{6}$}
\begin{minipage}[t]{14cm}
\textit{ The Henryk Niewodniczanski Institute of Nuclear Physics, Polish Academy of \\
Sciences, Krakow, Poland}

\end{minipage}

\makebox[3ex]{$^{7}$}
\begin{minipage}[t]{14cm}
\textit{ AGH University of Science and Technology, Faculty of Physics and Applied Computer
Science, Krakow, Poland}

\end{minipage}

\makebox[3ex]{$^{8}$}
\begin{minipage}[t]{14cm}
\textit{ Department of Physics, Jagellonian University, Krakow, Poland}

\end{minipage}

\makebox[3ex]{$^{9}$}
\begin{minipage}[t]{14cm}
\textit{ Deutsches Elektronen-Synchrotron DESY, Hamburg, Germany}

\end{minipage}

\makebox[3ex]{$^{10}$}
\begin{minipage}[t]{14cm}
\textit{ Deutsches Elektronen-Synchrotron DESY, Zeuthen, Germany}

\end{minipage}

\makebox[3ex]{$^{11}$}
\begin{minipage}[t]{14cm}
\textit{ School of Physics and Astronomy, University of Glasgow,
Glasgow, United Kingdom}~$^{D}$

\end{minipage}

\makebox[3ex]{$^{12}$}
\begin{minipage}[t]{14cm}
\textit{ GSI Helmholtzzentrum f\"{u}r Schwerionenforschung GmbH, Darmstadt, Germany}

\end{minipage}

\makebox[3ex]{$^{13}$}
\begin{minipage}[t]{14cm}
\textit{ Hamburg University, Institute of Experimental Physics, Hamburg,
Germany}~$^{E}$

\end{minipage}

\makebox[3ex]{$^{14}$}
\begin{minipage}[t]{14cm}
\textit{ Physikalisches Institut of the University of Heidelberg, Heidelberg, Germany}

\end{minipage}

\makebox[3ex]{$^{15}$}
\begin{minipage}[t]{14cm}
\textit{ Institute of Particle and Nuclear Studies, KEK,
Tsukuba, Japan}~$^{F}$

\end{minipage}

\makebox[3ex]{$^{16}$}
\begin{minipage}[t]{14cm}
\textit{ Institute of Physics and Technology of Ministry of Education and
Science of Kazakhstan, Almaty, Kazakhstan}

\end{minipage}

\makebox[3ex]{$^{17}$}
\begin{minipage}[t]{14cm}
\textit{ Institute for Nuclear Research, National Academy of Sciences, Kyiv, Ukraine}

\end{minipage}

\makebox[3ex]{$^{18}$}
\begin{minipage}[t]{14cm}
\textit{ Department of Nuclear Physics, National Taras Shevchenko University of Kyiv, Kyiv, Ukraine}

\end{minipage}

\makebox[3ex]{$^{19}$}
\begin{minipage}[t]{14cm}
\textit{ Meiji Gakuin University, Faculty of General Education,
Yokohama, Japan}~$^{F}$

\end{minipage}

\makebox[3ex]{$^{20}$}
\begin{minipage}[t]{14cm}
\textit{ Lomonosov Moscow State University, Skobeltsyn Institute of Nuclear Physics,
Moscow, Russia}~$^{G}$

\end{minipage}

\makebox[3ex]{$^{21}$}
\begin{minipage}[t]{14cm}
\textit{ Max-Planck-Institut f\"ur Physik, M\"unchen, Germany}

\end{minipage}

\makebox[3ex]{$^{22}$}
\begin{minipage}[t]{14cm}
\textit{ Department of Physics, University of Oxford,
Oxford, United Kingdom}~$^{D}$

\end{minipage}

\makebox[3ex]{$^{23}$}
\begin{minipage}[t]{14cm}
\textit{ INFN Padova, Padova, Italy}~$^{A}$

\end{minipage}

\makebox[3ex]{$^{24}$}
\begin{minipage}[t]{14cm}
\textit{ Dipartimento di Fisica e Astronomia dell' Universit\`a and INFN,
Padova, Italy}~$^{A}$

\end{minipage}

\makebox[3ex]{$^{25}$}
\begin{minipage}[t]{14cm}
\textit{ Polytechnic University, Tokyo, Japan}~$^{F}$

\end{minipage}

\makebox[3ex]{$^{26}$}
\begin{minipage}[t]{14cm}
\textit{ Raymond and Beverly Sackler Faculty of Exact Sciences, School of Physics, \\
Tel Aviv University, Tel Aviv, Israel}~$^{H}$

\end{minipage}

\makebox[3ex]{$^{27}$}
\begin{minipage}[t]{14cm}
\textit{ Department of Physics, Temple University, Philadelphia, PA 19122, USA}~$^{I}$

\end{minipage}

\makebox[3ex]{$^{28}$}
\begin{minipage}[t]{14cm}
\textit{ Department of Physics, Tokyo Institute of Technology,
Tokyo, Japan}~$^{F}$

\end{minipage}

\makebox[3ex]{$^{29}$}
\begin{minipage}[t]{14cm}
\textit{ Universit\`a di Torino and INFN, Torino, Italy}~$^{A}$

\end{minipage}

\makebox[3ex]{$^{30}$}
\begin{minipage}[t]{14cm}
\textit{ Universit\`a del Piemonte Orientale, Novara, and INFN, Torino,
Italy}~$^{A}$

\end{minipage}

\makebox[3ex]{$^{31}$}
\begin{minipage}[t]{14cm}
\textit{ Physics and Astronomy Department, University College London,
London, United Kingdom}~$^{D}$

\end{minipage}

\makebox[3ex]{$^{32}$}
\begin{minipage}[t]{14cm}
\textit{ Faculty of Physics, University of Warsaw, Warsaw, Poland}

\end{minipage}

\makebox[3ex]{$^{33}$}
\begin{minipage}[t]{14cm}
\textit{ National Centre for Nuclear Research, Warsaw, Poland}

\end{minipage}

\makebox[3ex]{$^{34}$}
\begin{minipage}[t]{14cm}
\textit{ Department of Particle Physics and Astrophysics, Weizmann
Institute, Rehovot, Israel}

\end{minipage}

\makebox[3ex]{$^{35}$}
\begin{minipage}[t]{14cm}
\textit{ Department of Physics, York University, Ontario, Canada M3J 1P3}~$^{J}$

\end{minipage}

}

\vspace{3em}

%  references concerning institutes;

{\setlength{\parskip}{0.4em}\raggedright
\makebox[3ex]{$^{ A}$}
\begin{minipage}[t]{14cm}
 supported by the Italian National Institute for Nuclear Physics (INFN) \
\end{minipage}

\makebox[3ex]{$^{ B}$}
\begin{minipage}[t]{14cm}
 supported by the German Federal Ministry for Education and Research (BMBF), under
 contract No.\ 05 H09PDF\
\end{minipage}

\makebox[3ex]{$^{ C}$}
\begin{minipage}[t]{14cm}
 supported by HIR grant UM.C/625/1/HIR/149 and UMRG grants RU006-2013, RP012A-13AFR and RP012B-13AFR from
 Universiti Malaya, and ERGS grant ER004-2012A from the Ministry of Education, Malaysia\
\end{minipage}

\makebox[3ex]{$^{ D}$}
\begin{minipage}[t]{14cm}
 supported by the Science and Technology Facilities Council, UK\
\end{minipage}

\makebox[3ex]{$^{ E}$}
\begin{minipage}[t]{14cm}
 supported by the German Federal Ministry for Education and Research (BMBF), under
 contract No.\ 05h09GUF, and the SFB 676 of the Deutsche Forschungsgemeinschaft (DFG) \
\end{minipage}

\makebox[3ex]{$^{ F}$}
\begin{minipage}[t]{14cm}
 supported by the Japanese Ministry of Education, Culture, Sports, Science and Technology
 (MEXT) and its grants for Scientific Research\
\end{minipage}

\makebox[3ex]{$^{ G}$}
\begin{minipage}[t]{14cm}
 partially supported by RF Presidential grant NSh-7989.2016.2\
\end{minipage}

\makebox[3ex]{$^{ H}$}
\begin{minipage}[t]{14cm}
 supported by the Israel Science Foundation\
\end{minipage}

\makebox[3ex]{$^{ I}$}
\begin{minipage}[t]{14cm}
 supported in part by the Office of Nuclear Physics within the U.S.\ DOE Office of Science
\end{minipage}

\makebox[3ex]{$^{ J}$}
\begin{minipage}[t]{14cm}
 supported by the Natural Sciences and Engineering Research Council of Canada (NSERC) \
\end{minipage}

}

\pagebreak[4]
{\setlength{\parskip}{0.4em}

%  references concerning members;

\makebox[3ex]{$^{ a}$}
\begin{minipage}[t]{14cm}
now at INFN Roma, Italy\
\end{minipage}

\makebox[3ex]{$^{ b}$}
\begin{minipage}[t]{14cm}
now at DST-Inspire Faculty, Department of Technology, SPPU, India\
\end{minipage}

\makebox[3ex]{$^{ c}$}
\begin{minipage}[t]{14cm}
now at Sant Longowal Institute of Engineering and Technology, Longowal, Punjab, India\
\end{minipage}

\makebox[3ex]{$^{ d}$}
\begin{minipage}[t]{14cm}
also at Agensi Nuklear Malaysia, 43000 Kajang, Bangi, Malaysia\
\end{minipage}

\makebox[3ex]{$^{ e}$}
\begin{minipage}[t]{14cm}
supported by the Polish National Science Centre (NCN) grant no. DEC-2014/13/B/ST2/02486\
\end{minipage}

\makebox[3ex]{$^{ f}$}
\begin{minipage}[t]{14cm}
now at Rockefeller University, New York, NY 10065, USA\
\end{minipage}

\makebox[3ex]{$^{ g}$}
\begin{minipage}[t]{14cm}
now at University of Liverpool, United Kingdom\
\end{minipage}

\makebox[3ex]{$^{ h}$}
\begin{minipage}[t]{14cm}
now at Tel Aviv University, Israel\
\end{minipage}

\makebox[3ex]{$^{ i}$}
\begin{minipage}[t]{14cm}
now at Physikalisches Institut, University of Heidelberg, Germany\
\end{minipage}

\makebox[3ex]{$^{ j}$}
\begin{minipage}[t]{14cm}
also at Physikalisches Institut of the University of Heidelberg, Heidelberg,  Germany\
\end{minipage}

\makebox[3ex]{$^{ k}$}
\begin{minipage}[t]{14cm}
Alexander von Humboldt Professor; also at DESY and University of Oxford\
\end{minipage}

\makebox[3ex]{$^{ l}$}
\begin{minipage}[t]{14cm}
also at DESY\
\end{minipage}

\makebox[3ex]{$^{ m}$}
\begin{minipage}[t]{14cm}
now at European X-ray Free-Electron Laser facility GmbH, Hamburg, Germany\
\end{minipage}

\makebox[3ex]{$^{ n}$}
\begin{minipage}[t]{14cm}
also at GSI Helmholtzzentrum f\"{u}r Schwerionenforschung GmbH, Darmstadt, Germany\
\end{minipage}

\makebox[3ex]{$^{ o}$}
\begin{minipage}[t]{14cm}
now at Kobe University, Japan\
\end{minipage}

\makebox[3ex]{$^{ p}$}
\begin{minipage}[t]{14cm}
now at RWTH Aachen, Germany\
\end{minipage}

\makebox[3ex]{$^{ q}$}
\begin{minipage}[t]{14cm}
also at University of Bristol, United Kingdom\
\end{minipage}

\makebox[3ex]{$^{ r}$}
\begin{minipage}[t]{14cm}
also at Max Planck Institute for Physics, Munich, Germany, External Scientific Member\
\end{minipage}

\makebox[3ex]{$^{ s}$}
\begin{minipage}[t]{14cm}
now at University of Bath, United Kingdom\
\end{minipage}

\makebox[3ex]{$^{ t}$}
\begin{minipage}[t]{14cm}
also supported by DESY\
\end{minipage}

\makebox[3ex]{$^{ u}$}
\begin{minipage}[t]{14cm}
also at Lodz University, Poland\
\end{minipage}

\makebox[3ex]{$^{ v}$}
\begin{minipage}[t]{14cm}
now at Polish Air Force Academy in Deblin\
\end{minipage}

}

}

\cleardoublepage
%
%------------------------------------------------------------------------------
%
\pagenumbering{arabic}
%
% Comment out this line to remove date/time for final version
%
% \pagestyle{scrheadings}
%------------------------------------------------------------------------------
%       Text
%------------------------------------------------------------------------------
% -*- mode: LaTeX; mode: flyspell; -*-

% ----------------------------------------------------------------------------
%       Introduction
% ----------------------------------------------------------------------------

\section{Introduction}
\label{sec-int}

The H1 and ZEUS collaborations measured inclusive 
$e^{\pm}p$ scattering cross sections at HERA from
1994 to 2000 (HERA I) and from 2002 to 2007 (HERA II), 
collecting a total integrated luminosity of approximately 1\,fb$^{-1}$.
All the inclusive data sets were combined \cite{h1zeus_inc} to create
one consistent set of neutral current (NC) and charged current (CC)
cross-section measurements for $e^{\pm}p$ scattering with unpolarised beams.
These cross sections were used as input to a QCD analysis
within the
DGLAP~\cite{Gribov:1972ri,Gribov:1972rt,Lipatov:1974qm,Dokshitzer:1977sg,Altarelli:1977zs} 
formalism of the Standard Model (SM),
resulting in parton distribution function (PDF)
parameterisations of the proton denoted as \mbox{HERAPDF2.0} \cite{h1zeus_inc}.
Precise knowledge of the parton densities inside the proton is
crucial, in particular, for 
the full exploitation of the physics potential of the LHC.

HERA measurements of deep inelastic $e^\pm p$ scattering (DIS)
cross sections at the highest values of negative
four-momentum-transfer squared,  $Q^2$, can be sensitive to
beyond the Standard Model (BSM)
contributions even at scales far beyond the centre-of-mass energy 
of 320\,GeV.
For many ``new physics'' scenarios, cross sections can be affected
by new kinds of interactions in which virtual BSM particles are exchanged.
As the HERA centre-of-mass energy is assumed to be far below the scale of
the new physics, all such BSM interactions can be approximated as
contact interactions (CIs).

In the absence of direct observation of BSM physics at HERA, the ZEUS
collaboration has used the HERA combined measurement 
of inclusive cross sections  \cite{h1zeus_inc} to set limits on possible
deviations from the SM due to a finite quark radius
\cite{rq_paper}. 
 If BSM physics effects existed in the HERA data, the current PDF
 sets would have been biased by partially or totally absorbing
 unrecognised BSM contributions. A new approach was therefore used to
 minimise this bias, based on simultaneous fits of the PDFs and the
 contributions of ``new physics'' processes.  
It was assumed,
as usual, that the use of SM Monte Carlo
simulations in the original extraction of the cross sections did not
introduce a significant bias.
The new procedure \cite{rq_paper} introduced to set limits on the
quark radius has also been extended to other ``new physics'' scenarios.
BSM contributions were added to the fit to HERA data to
investigate whether this results in an improvement of the description
of the data.
Previous CI searches \cite{Chekanov:2003pw,Aaron:2011mv,Collaboration:2011qaa}
used only a subset of the HERA data used here.

%
% ----------------------------------------------------------------------------
%  Models for new physics
% ----------------------------------------------------------------------------
%

\section{Models for new physics}
\label{sec:models}

Four-fermion CIs represent an effective theory 
which describes low-energy effects  due to physics at much higher 
energy scales.
Contact-interaction models can describe the effects of heavy leptoquarks,
additional heavy weak bosons and electron or quark
compositeness~\cite{Tanabashi:2018oca}. 
The CI approach is not renormalisable 
and is only valid in the low-energy limit, far below the mass scale of
the new physics.
For HERA data collected at centre-of-mass energy of
  320\,GeV this approach should be applicable for
  new physics mass scales of about 1\,TeV or above. 
Vector CI currents considered here
are represented by additional terms in the SM Lagrangian:
\begin{eqnarray}
{\cal L}_\text{CI} & = & 
   \sum_{^{i,j=L,R}_{q=u,d,s,c,b,t}} 
   \eta^{eq}_{ij} (\bar{e}_{i} \gamma^{\mu} e_{i} )
                 (\bar{q}_{j} \gamma_{\mu} q_{j}) \; ,
	\label{eq-cilagr}
\end{eqnarray}
where the sum runs over electron and quark chiralities
and quark flavours.
The couplings  $\eta^{eq}_{ij}$ describe 
the chiral and flavour structure of CIs.

The CI contributions to the electron--quark scattering amplitude do
not depend on the $Q^{2}$ scale of the process.
Thus, the CI contributions to the DIS cross sections are expected
to be largest, relative to the SM contribution, at the highest $Q^2$
values~\cite{pl:b591:23}. 
However, the exact shape of the expected deviations from the SM predictions
depends on the assumed flavour and chiral structure of the CI couplings.
Depending on the coupling structure, different
contributions are also expected for
electron--proton and positron--proton scattering.

% ------------- CI  ---------------

\subsection{Contact interactions}
\label{sec:ci}

For this study, the CI scenarios were defined assuming that all quarks
have the same CI couplings: 
\begin{eqnarray}
\eta^{eu}_{ij} \; = \; \eta^{ed}_{ij} & = \; \eta^{es}_{ij} & 
= \; \eta^{ec}_{ij} \; = \; \eta^{eb}_{ij} \; = \; \eta^{et}_{ij} \;, \nonumber
\end{eqnarray}
leading to four independent couplings, $\eta^{eq}_{ij}$, with $i,j=L,R$. 
Owing to the impracticality of setting limits in a four-dimensional
parameter space, a set of one-parameter scenarios was analysed. 
Each scenario is defined by a set of four coefficients,
$\epsilon_{ij}$, each of which may take the values $\pm1$ or
zero, see \tab{cifit}, and by the coupling strength $\eta$ or compositeness
scale $\Lambda$.
The couplings are given by the formula
\begin{eqnarray}
  \eta^{eq}_{ij} & = &
  \eta \; \epsilon_{ij} \;  = \;
  \pm \frac{4 \pi}{\Lambda^{2}} \; \epsilon_{ij} \;  . \nonumber
\end{eqnarray}
Four parity-violating scenarios were selected for this study. They
are listed in the upper part of \tab{cifit}. In addition, nine
scenarios conserving parity were chosen, shown in the lower part of
\tab{cifit}, for which the bounds resulting from atomic-parity
violation measurements are not relevant. 
These models were first introduced in a previous ZEUS
publication~\cite{Breitweg:1999ssa}.
Note that the coupling strength $\eta$ can be both positive and negative, 
and the two cases are distinct because of the interference with the SM
amplitudes~\cite{Tanabashi:2018oca,pl:b591:23}.
Only in the case of the VA model is the contribution of the
interference term negligible, so that the model predictions
are mainly sensitive to $\eta^2$. 
When setting limits for BSM contributions, scenarios with positive and negative
$\eta$ values were considered separately.

%-------------- LQ -----------------

\subsection{Heavy leptoquarks}
\label{sec:lq}

Leptoquarks (LQs) appear in certain extensions of the SM that 
connect leptons and quarks; they carry both lepton and 
baryon numbers and have spin 0 (in case of scalar LQs) or 1 (vector LQs). 
According to the general classification 
proposed by Buchm\"uller, R\"uckl and Wyler \cite{pl:b191:442e},
there are 14 possible LQ types (isospin singlets or multiplets):
seven scalar and seven vector\footnote{Leptoquark states are named
according to the so-called Aachen notation \cite{zfp:c46:679}.}.
In the limit of heavy LQs (for masses much higher than the HERA centre-of-mass energy,
$M_\text{LQ} \gg \sqrt{s}$),
the effect of $s$- and $t$-channel LQ exchange
is equivalent to a vector-type $eeqq$ contact interaction\footnote{%
For the invariant mass range accessible at HERA, with $\sqrt{s}\approx 320\gev$,
the heavy LQ approximation is already applicable for $M_\text{LQ} >
400\gev$ \cite{Abramowicz:2012tg}.
This condition is fulfilled for all scenarios considered here
  unless $\lambda_\text{LQ} \ll 1$.
}. 
The effective LQ coupling, $\eta_\textrm{LQ}$, is given by
the square of the ratio of the leptoquark Yukawa 
coupling, $\lambda_\text{LQ}$, to the leptoquark mass, $M_\text{LQ}$:
\begin{eqnarray}
\eta_\textrm{LQ} & = & 
   \left(\frac{\lambda_\text{LQ}}{M_\text{LQ}} \right)^{2} \; . \nonumber
\end{eqnarray}
The CI couplings of the Lagrangian (Eq.~\ref{eq-cilagr}), $\eta^{eq}_{ij}$,
can be then written as
\begin{eqnarray}
\eta^{eq}_{ij} & = & 
   a^{eq}_{ij}\cdot \eta_\textrm{LQ} \; = \; 
   a^{eq}_{ij}\left(\frac{\lambda_\text{LQ}}{M_\text{LQ}} \right)^{2} \; , \nonumber
\end{eqnarray}
where the coefficients $a^{eq}_{ij}$ depend on the LQ species
\cite{zfp:c74:595} and are twice as large for vector as for 
scalar leptoquarks. 
By definition, the values of $\eta_\textrm{LQ}$ are
positive.\footnote{Note that five scalar and five vector LQ models  
correspond to the same coupling structure but with the opposite
coupling sign for scalar and vector scenarios.}
In the analysis presented in this paper,
leptoquark couplings are assumed to be family diagonal and
only the first-generation LQs are considered, $q = u, d$.  
Mass degeneration is assumed for leptoquark states within isospin
  doublets and triplets.
Contrary to the considered CI scenarios, the LQ coupling structure is
different for $u$ and $d$ quarks, resulting in different shapes of
the expected cross-section deviations. 
The coupling structure for different leptoquark species is shown in \tab{lqfit}.

% ----------------------------------------------------------------------------
%       QCD+CI analysis
% ----------------------------------------------------------------------------

\section{Extended fit to the inclusive HERA data}
\label{sec-fit}

\subsection{QCD+CI fit procedure}

The analysis is based on a comparison of the measured inclusive cross
sections with the model predictions.
The effects of each CI scenario are taken into account by 
scaling the NLO QCD predictions at given values of $x$ and $Q^{2}$,
corresponding to the inclusive cross-section
measurements~\cite{h1zeus_inc}, with the cross-section ratio
\begin{eqnarray}
  R_\text{CI} & = & \begin{array}{c}
               \frac{~~d^2\sigma}{dx\,dQ^2} ^\text{SM+CI} \\[1mm] \hline \\[-5.5mm]
               \frac{~~d^2\sigma}{dx\,dQ^2}  ^\text{SM} 
               \end{array}
                                                       \label{eq-rci}
\end{eqnarray}
calculated in leading order in electroweak and CI couplings.

The QCD analysis presented in this paper follows the approach adopted for
the determination of HERAPDF2.0~\cite{h1zeus_inc}.
This analysis is extended to take into account the possible BSM contributions
to the expected cross-section values, as described previously~\cite{rq_paper}.
The PDFs of the proton are described
at a starting scale of $1.9$ GeV$^2$ in terms of 14 parameters.  
These parameters, denoted $p_k$ in the following (or $\boldsymbol{p}$
for the set of parameters), together with the possible contribution of
BSM phenomena (described by the CI coupling $\eta$) were
fitted to the data using a $\chi^2$ method, with the $\chi^2$ formula
given by
\begin{equation}
 \chi^2 \left(\boldsymbol{p},\boldsymbol{s},\eta \right) = %\\
%~~~=
 \sum_i
 \frac{\left[m^i
+ \sum_j \gamma^i_j m^i s_j  - {\mu_{0}^i} \right]^2}
{\left( \textstyle \delta^2_{i,\textrm{stat}} +
\delta^2_{i,\textrm{uncor}} \right) \,  (\mu_{0}^i)^2}
 + \sum_j s^2_j ~~.
\label{eq:qcdfit}
\end{equation} 
Here, $\mu_{0}^{i}$ and $m^i$ are, respectively, the measured
cross-section values and the SM+CI cross-section predictions at the
point $i$. 
The quantities $\gamma^{i}_j $, $\delta_{i,\textrm{stat}} $ and 
$\delta_{i,\textrm{uncor}}$ are, respectively, the relative correlated 
systematic, the relative statistical and the relative uncorrelated 
systematic uncertainties on the input data. 
The components $s_j$ of the vector $\boldsymbol{s}$ are the
correlated systematic shifts of the cross sections (given in units of
the respective correlated systematic uncertainties), which were fitted 
to the data together with PDF parameter set
$\boldsymbol{p}$ and the CI coupling $\eta$. 
The summations extend over all data points $i$
and all correlated systematic uncertainties $j$.

All fits presented here were performed within the xFitter
framework~\cite{xFitter} modified to include CI contributions. 
The $\chi^2$ formula was different with respect to that used for
the \mbox{HERAPDF2.0} study~\cite{h1zeus_inc} to reflect
the fact that fixed Gaussian uncertainties on the input data
points were assumed.  
The same assumption was also used when generating the data replicas,
see Section~\ref{sec-limit}. 

When not taking into account the CI contribution, the resulting sets of
PDFs, referred to as ZCIPDFs in the following, 
are in good agreement with \mbox{HERAPDF2.0} fit results 
obtained within the HERAFitter framework~\cite{HERAFitter}.
The experimental uncertainties on the fitted model parameters 
and on the predictions from ZCIPDF, resulting from the uncertainties 
of the input HERA data, were defined by the criterion $\Delta\chi^2=1$. 
This takes into account statistical uncertainties and also correlated
and uncorrelated systematic uncertainties of the combined HERA data,
see Eq.~\ref{eq:qcdfit}.

\subsection{Modelling uncertainties}
\label{sec:assumpt}

Following the approach used for the \mbox{HERAPDF2.0} fit~\cite{h1zeus_inc},
the uncertainties on the ZCIPDF fit due to the choice of the form of the 
parameterisation and the model settings were evaluated by  varying the
assumptions.  
Two kinds of parameterisation uncertainties were considered:
the variation in the fit starting scale, $\mu^2_{f_{0}}$,
and the addition of parameters in the parton-density parameterisation.
The parameters $D$ and $E$, defined in the previous HERA
analysis~\cite{h1zeus_inc}, were added separately for each PDF.
The final parameterisation uncertainty on the fitted coupling constant
was taken to be the largest of the resulting deviations.
The variations of charm and beauty mass parameters, $M_c$ and $M_b$, respectively,
were chosen in accordance with the mass estimation from HERAPDF2.0.
The variation of the strange-sea fraction, $f_s$, was chosen to span the ranges
between a suppressed strange sea~\cite{Martin:2009iq,Nadolsky:2008zw} and
an unsuppressed strange sea~\cite{Aad:2012sb,Aaboud:2016btc}. 
In addition to these model variations, the minimal $Q^2$ of the data points used in the
fit, $Q^2_\text{min}$,  was varied.
A summary of the variations on the model settings is given in \tab{model}. 

For each fitted coupling, the differences between the central fit and
the fits corresponding to the variations of $Q^2_\text{min}$, $f_s$,
$M_c$ and $M_b$, and the largest parameterisation uncertainty were
added in quadrature, separately for positive and negative deviations,
and represent the modelling uncertainty of the fit.  
The total uncertainty was obtained by adding in quadrature 
the experimental and the modelling uncertainties.

\subsection{Fit results}
\label{sec-fitres}

The ZCIPDF fit to the HERA inclusive data was extended by adding the
CI coupling, $\eta$ (or $\eta_\text{LQ}$ for  LQ models) as an additional fit parameter.
Results of the simultaneous QCD+CI fit to the HERA inclusive data, in terms of 
the fitted coupling values, are presented in \tab{cifit} and \tab{lqfit}, 
for  CI and heavy-LQ scenarios, respectively.
Experimental, modelling and total uncertainties on the fitted coupling values 
were calculated following the HERAPDF2.0 approach, as described above.
Also shown is the change of the $\chi^2$ value from the nominal SM fit,
$\Delta\chi^2=\chi^2_\textrm{SM+CI}-\chi^2_\textrm{SM}$. 
For most of the considered CI scenarios, only one minimum was observed 
in the $\chi^2$ dependence on the coupling value.
The VA model was the only case where two minima were observed, 
one for a positive and one for a negative coupling.
Results for both minima are presented in \tab{cifit}.

In most cases, correlations between PDF parameters and CI coupling values 
resulting from the QCD+CI fit are small. 
The largest correlations are observed  between the CI coupling and the
parameters $B_{d_{v}}$, $B_{u_{v}}$ and $C_{\bar{D}}$, used in the
description of the valence $d$ quark, valence $u$ quark and  
$d$-type anti-quark distribution, respectively.
Their absolute values reach 0.61 for  CI models 
($\eta - B_{d_{v}}$ correlation in the AA model) and 0.57 for LQ
models ($\eta - C_{\bar{D}}$ correlation in the $V_{0}^{L}$ model).

\begin{figure}[tbp]
\begin{center}
\includegraphics[width=0.9\textwidth]{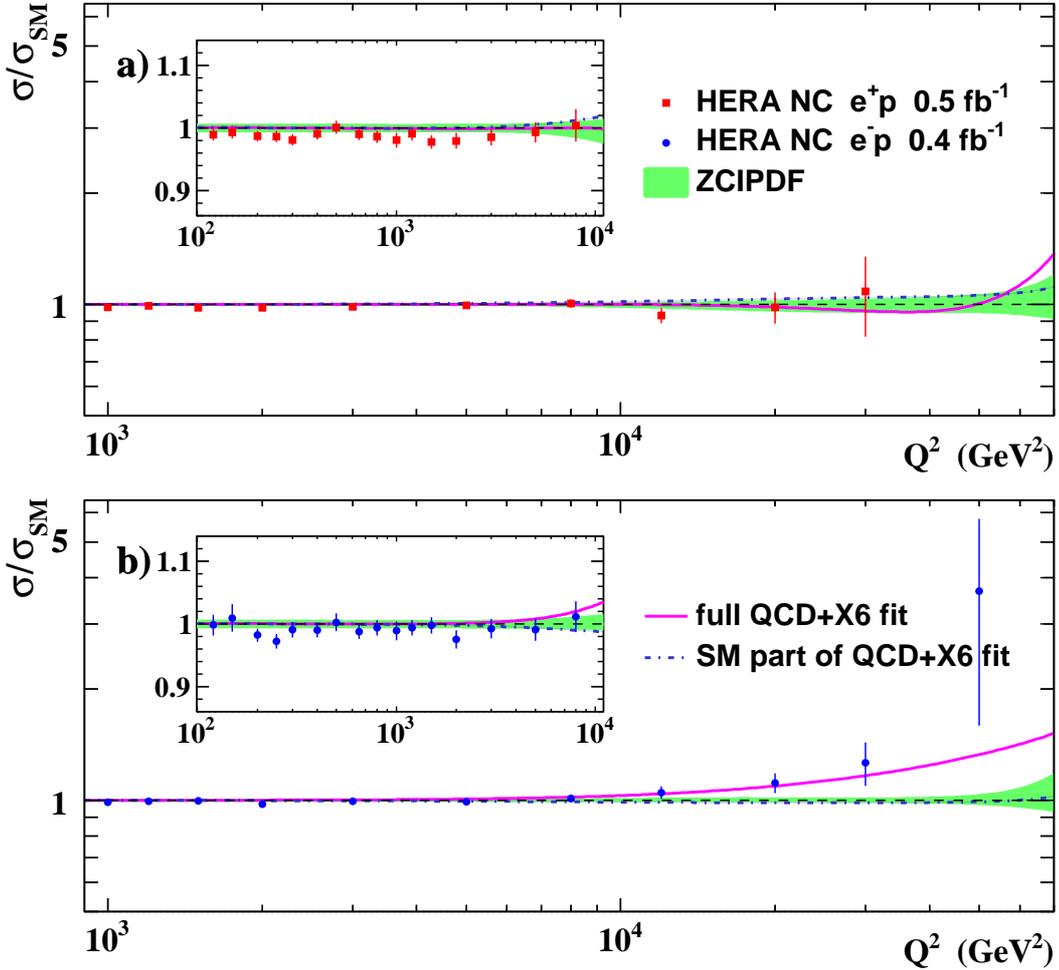}
\end{center}
  \caption{
Result of the simultaneous QCD+CI fit to the HERA inclusive data,
for the X6 CI model, compared to the combined HERA (a) $e^+p$ and
(b) $e^-p$ NC DIS data, relative to the SM expectations based on
the QCD fit without the CI contribution (ZCIPDF).
The bands represent the total uncertainty of the SM expectations. 
Also shown is the SM contribution to the cross section resulting
from the combined fit.
         }
  \label{fig-x6}
\end{figure}

\begin{figure}[tbp]
\begin{center}
\includegraphics[width=0.9\textwidth]{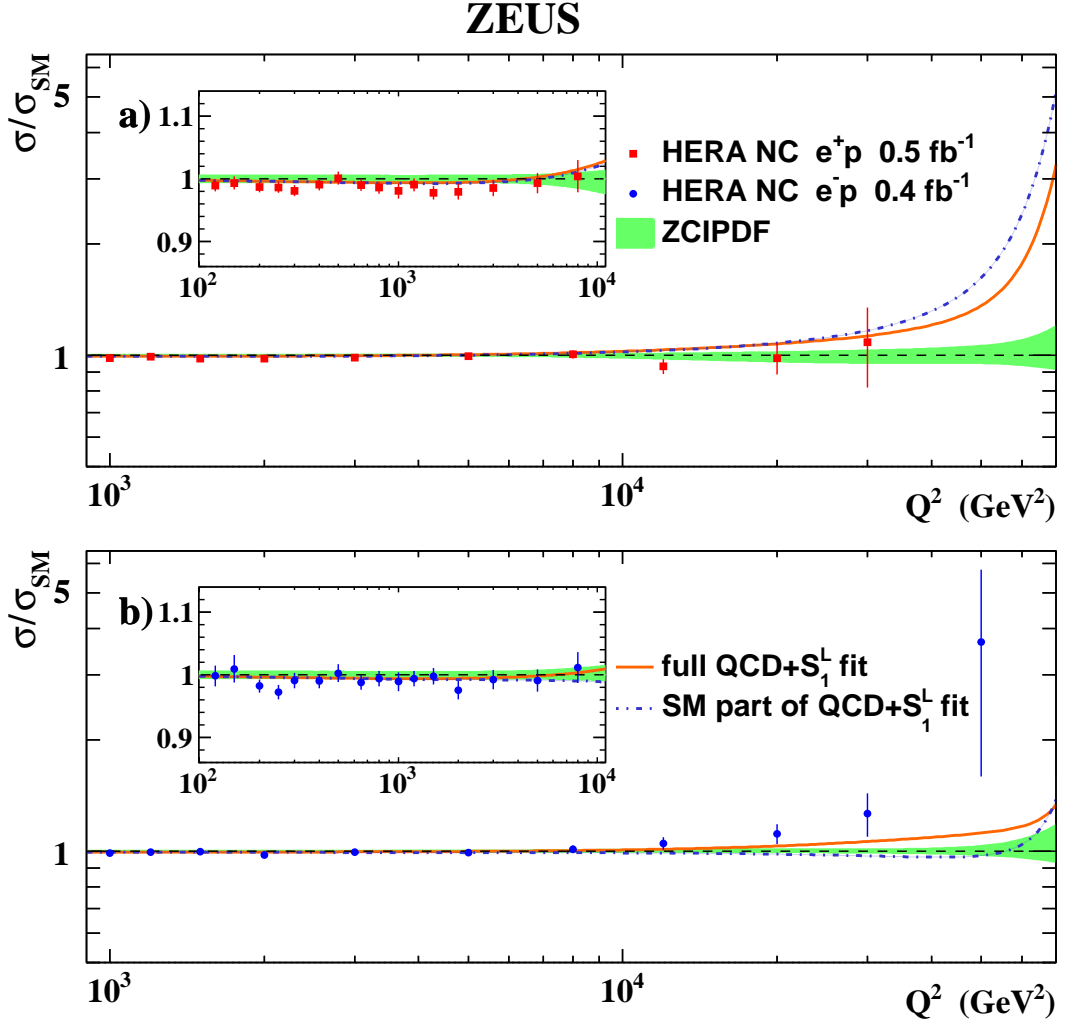}
\end{center}
  \caption{
Result of the simultaneous QCD+LQ fit to the HERA inclusive data,
for the $S_1^L$ LQ model in the contact interaction limit,
$M_{LQ}\gg \sqrt{s}$, compared to the combined HERA (a) $e^+p$ and
(b) $e^-p$ NC DIS data, relative to the SM expectations based on the
QCD fit without the CI contribution (ZCIPDF).
The bands represent the total uncertainty of the SM expectations. 
Also shown is the SM contribution to the cross section resulting
from the combined fit.
         }
  \label{fig-s1l}
\end{figure}

For six out of 13 considered  CI scenarios and seven out of 14
heavy-LQ models, no significant improvement in description of the data
was observed.
The fits were consistent with $\Delta\chi^2 \approx -1$, expected for
a reduction of the number of degrees of freedom. The fitted coupling
values for these models are consistent  with zero.
However, there are also four models (three CI and one\footnote{The improvement
observed for the $V_{0}^{L}$ model is not considered, as it is obtained for 
an unphysical (negative) coupling value.} LQ scenario), 
which result in an improved description of the data, 
with $\Delta\chi^2 < -4$.
The best description of the inclusive HERA data is obtained for the X6 
model ($\Delta\chi^2 = -6.01$) and $S_1^L$ model ($\Delta\chi^2 = -11.10$). 
The fit results for these models are compared with HERA NC DIS data in
\fig{x6} and \fig{s1l}, respectively.
Also indicated is the SM contribution to the NC DIS cross sections obtained 
from the QCD+CI fit.

Figure \ref{fig-x6} shows that,
for the X6 model, the determination of the proton PDFs is affected
very little by the CI contribution; the SM part of the NC DIS cross
sections extracted from the QCD+CI fit agree with the nominal ZCIPDF
fit within the quoted PDF uncertainties.  
The change in the predicted NC DIS cross section is dominated by the
CI contribution. 
The situation is different for the $S_1^L$ heavy-LQ model 
shown in \fig{s1l}, where
the description of the proton PDFs is significantly affected 
when the heavy-LQ contribution is taken into account in the fit. 
As a result, the cross-section prediction for NC $e^+ p$ DIS due to
$\gamma/Z^0$ exchange increases at the highest values of $Q^2$,
$Q^2>50\,000$\,GeV$^{2}$, by about a factor of two. 
The virtual leptoquark exchange contribution to the NC $e^+ p$ DIS
cross section is much smaller than the change observed in the SM
contribution. 
Moreover, it decreases the total cross section due to the negative
interference with the SM part.
The improvement in the description of the data for the $S_1^L$ heavy-LQ
model is also due to the better agreement of the resulting
predictions with the CC DIS data.

\tab{cifit} also includes estimates of the modelling 
(and the resulting total) uncertainties on the fitted CI coupling values.
For most of the models, the modelling uncertainties tend to be small, below
the level of experimental uncertainties. 
However, they are significant for the three CI models, AA, X1 and X6,
for which $\Delta\chi^2 < -4$.
Most important are the choice of the $Q^{2}_\textrm{min}$ parameter
(used to select the input data set for the fit) and the inclusion of the
additional  $D_{u_{v}}$ parameter \cite{h1zeus_inc} in the valence
$u$-quark density description.  
When the $D_{u_{v}}$ parameter is added to the ZCIPDF,
it results in a $\Delta\chi^2 = -10.3$.
After the addition of the $D_{u_{v}}$ parameter, the improvement in
$\chi^2$ due to the CI term is reduced for the AA, X1 and X6 models,
and it is maximal for the VA model at $\Delta\chi^2 = -4.6$.

For the $S_{1}^{L}$ LQ scenario, although modelling uncertainties are
sizeable, see \tab{lqfit}, they cannot explain the improvement in the
description of the HERA data. 
The QCD+$S_{1}^{L}$ fits result in $\Delta\chi^2 < -9$ for all
considered model and parameterisation variations.
The  $S_{1}^{L}$ contribution to the predicted NC $e^- p$ DIS cross
section increases at the highest $Q^2$ values.
For $e^+ p$ scattering, the increase is expected at large $x$
values\footnote{This is due to an additional kinematic factor of
$(1-y)^2$ multiplying the LL scattering amplitude for NC $e^+ p$
DIS.}.   
A cross-check was also made using the bilog parameterisation~\cite{xFitter}.
While the overall description achieved with this quite different ansatz
is much worse than the description achieved with the
HERAPDF parameterisation, an $S_{1}^{L}$ term of similar strength was found.
%

% ----------------------------------------------------------------------------
%       Statistical analysis
% ----------------------------------------------------------------------------

\section{Limit-setting precedure}
\label{sec-limit}

The limits on the mass scales of the CI and heavy-LQ models were
derived  in a frequentist approach \cite{Cousins:1994yw} using the
technique of replicas. 

\subsection{Data set replicas}

Replicas are sets of cross-section values,
corresponding to the HERA inclusive data set
that are generated by varying all cross sections randomly according to their
known uncertainties.
For the analysis presented here, multiple replica sets were used, each 
covering cross-section values on all points of the $x,Q^2$ grid
used in the QCD fit.
For assumed true values of the CI coupling,
$\eta^\textrm{True}$, replica data sets were created by taking
the reduced cross sections calculated from the nominal PDF fit
(with CI coupling $\eta \equiv 0$) and scaling them
with the cross-section ratio $R^\text{CI}$ given by Eq.~\ref{eq-rci}.
This results in a set of cross-section values $m_{0}^{i}$ for the assumed
true CI coupling  $\eta^\textrm{True}$.
The values of $m_{0}^{i}$ were then varied randomly within the statistical
and systematic uncertainties taken from the data, taking correlations
of systematic uncertainties into account.
All uncertainties were assumed to follow a Gaussian
distribution\footnote{It was verified that using a Poisson probability
distribution for producing replicas at high $Q^2$, where the event
samples are small, and using the $\chi^2$ minimisation for
these data did not change the results.}.
For each replica, the generated value of the cross section 
at the point $i$, $\mu^{i}$, was calculated as
\begin{equation}
 \mu^{i}  = 
 \left[ m_{0}^{i} + \sqrt{\delta^2_{i,\textrm{stat}} + \delta^2_{i, \textrm{uncor}}} \cdot  \mu_{0}^{i} \cdot r_i \right]
\cdot
\left( 1 + \sum_j \gamma^i_j \cdot r_j  \right)~~,
\label{eq:replica}
\end{equation} 
where variables $r_i$ and $r_j$ represent random numbers from a normal
distribution for each data point $i$ and for each source of correlated
systematic uncertainty $j$, respectively.

The approach adopted was to generate large sets of replicas and  use them
to test the hypothesis that the cross sections are consistent with
the SM predictions or that they were modified by a fixed
CI coupling according to Eq.~\ref{eq-rci}.
The fitted $\eta$ values from the replicas, $\eta^\textrm{Fit}$,
were used as a test statistics and compared to the corresponding value
$\eta^\textrm{Data}$  determined from a fit to the data.

To quantify the statistical consistency of the fit results with the SM
expectations, the probability that an experiment assuming the validity
of the SM (replicas generated with
$\eta^\textrm{True} = \eta^\textrm{SM} \equiv 0$) would produce a
value of $\eta^\textrm{Fit}$ greater than (or less than) 
that obtained from the data was calculated:
\begin{equation}
p_\text{SM} = \left\{
\begin{array}{lll}
p(\eta^\textrm{Fit} > \eta^\textrm{Data}) & \textrm{for} & \eta^\textrm{Data}>0 \;, \\
p(\eta^\textrm{Fit} < \eta^\textrm{Data}) &  \textrm{for} & \eta^\textrm{Data}<0 \;,
\end{array}
\right. \label{eq:psm}
\end{equation}
where the probability $p$ was calculated from the distribution of
$\eta^\textrm{Fit}$ values for a large set of generated SM replicas.

\subsection{Constraining BSM scenarios}

While for LQ models only positive $\eta_\text{LQ}$ values were considered, 
in the case of the CI scenarios, coupling limits were calculated
separately for positive and negative $\eta$ values. 
The upper (lower) \CL{95} limit on the positive (negative) coupling,
$\eta^+$ ($\eta^-$), for a given scenario was determined as the value
of $\eta^\textrm{True}$ for which 95\% of the replicas produced a
fitted coupling value, $\eta^\textrm{Fit}$, larger (smaller) than that
found in the data, $\eta^\textrm{Data}$, see Eq.~\ref{eq:psm}. 
The corresponding mass-scale values ($\Lambda^+$ and $\Lambda^-$ for
CI scenarios or $M/\lambda$ for LQ models) will be referred to as
mass-scale limits.  
A similar procedure~\cite{rq_paper} was also used to calculate the
expected limit values, which were defined by comparing replica fit
results with $\eta^\textrm{SM} \equiv 0$ instead of $\eta^\textrm{Data}$. 

To take modelling uncertainties into account, the limit-calculation 
procedure was repeated for model or parameterisation variations resulting 
in the highest and the lowest $\eta^\textrm{Data}$ values for each model. 
The weakest of the obtained coupling limits was taken as the result 
of the analysis and used to calculate the final mass-scale limits.
This is clearly the most conservative approach, which is motivated by
the difficulty in defining the underlying probability distribution for
some of the considered modelling variations.

The expected limits are not sensitive to the modelling variations
because these mainly affect the data fit results ($\eta^\textrm{Data}$
values) and the expected values do not depend on $\eta^\textrm{Data}$
(replica fit results are compared to $\eta^\text{SM} \equiv 0$).
Therefore, these variations were not considered for the expected limits.

For each CI and LQ scenario, at least 3000 Monte Carlo replicas were
generated and fitted for each value of $\eta^\textrm{True}$.
When using xFitter to perform replica fits, the
  inclusion of more models in the analysis was limited by the
  processing time. 
To facilitate efficient processing of replica data,
a simplified fit method, based on the Taylor
expansion of the cross-section predictions in terms of PDF
parameters was developed, which reduced the processing time
by almost two orders of magnitude \cite{ci_note}.

% ----------------------------------------------------------------------------
%       Results
% ----------------------------------------------------------------------------

\section{Results}
\label{sec:res}

% ---------- CI ----------------

The probabilities $p_\text{SM}$ (Eq.~\ref{eq:psm}) calculated for
the considered CI scenarios with the SM replica sets are presented in
\tab{ci}. 
The statistical approach based on Monte Carlo replicas confirms the
observations described in Section~\ref{sec-fitres}, based on the
$\Delta\chi^2$ values. 
For six  CI models (LR, RL, VV, X2, X4 and X5), $p_\text{SM}$ is above 20\%, 
corresponding to less than a $1\,\sigma$ deviation from the nominal fit result 
($\eta^\textrm{Fit} = \eta^\textrm{SM} \equiv 0$).
For four models (LL, LR, VA and X3), the data fit results are 
reproduced by the SM replicas with 3--7\% probability, corresponding 
to about a $2\,\sigma$ difference.
However, for the three scenarios (AA, X1 and X6) with $\Delta\chi^2<-4$, 
$p_\text{SM}$ is below 1\%.
This confirms that the differences between the HERA data and the SM
predictions described by the additional CI contribution in the fit are
unlikely to be due to statistical fluctuations only.
As already discussed in Section \ref{sec-fitres}, the effect can be explained 
to some extent by the modelling uncertainties, in particular by the
deficiencies in the functional form used for the PDF parameterisation.

\begin{figure}[tbp]
\begin{center}
  \includegraphics[width=0.8\textwidth]{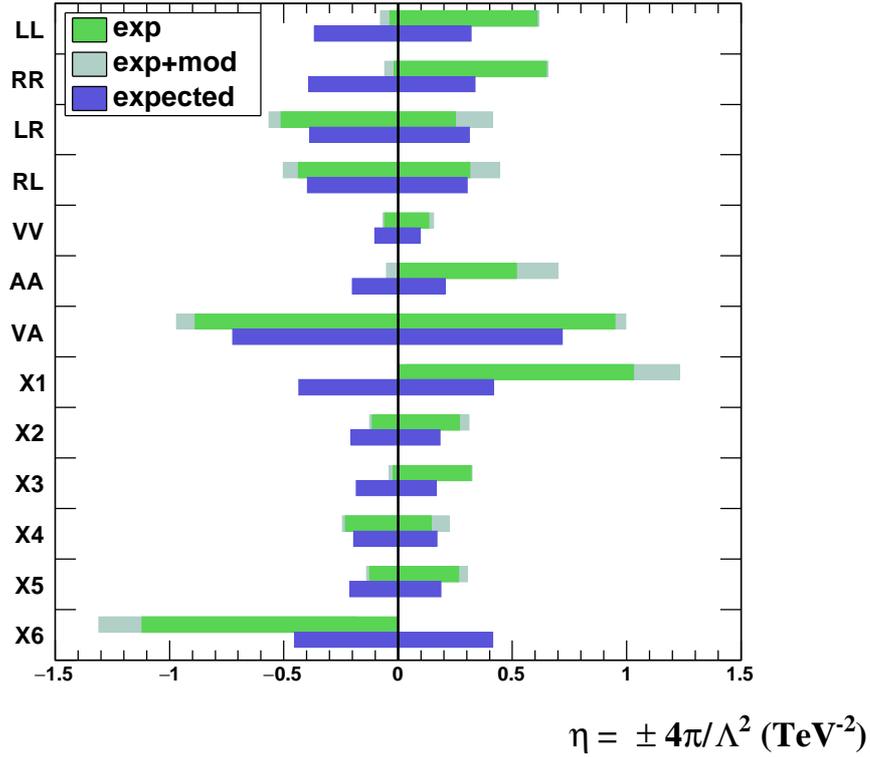}
\end{center}
  \caption{
    Limits on the CI coupling strength, $\eta =\pm 4 \pi /\Lambda^2$,
    evaluated  at \CL{95}. Compared are the limits
    calculated without (end points of dark upper bars) and with
    (light upper bars) modelling uncertainties,
    and the expected limits (lower bars).
    Limits are calculated separately for positive
    and negative coupling values (see text for details).
    }
  \label{fig-bars}
\end{figure}

Also shown in \tab{ci} are the \CL{95} limits on the coupling values,
$\eta^-$ and $\eta^+$, for different CI models.
Limits calculated without (exp) and with (exp+mod) model and
parameterisation variations, as described above, are compared to the
expected coupling limits in \fig{bars}. 
Coupling limits can also be translated into limits on the compositeness 
scales for the considered CI scenarios, also included in  \tab{ci}.

For most of the CI scenarios considered, the interference term gives a
significant contribution to the cross section and the sign of the CI
coupling is well constrained in the fit. 
However, in the case of the VA model, the contribution from the
interference term is much smaller than the direct CI contribution,
which is proportional to the coupling squared.
As a result, the model predictions are hardly sensitive to the
coupling sign and the global minimum of the $\chi^2$ function is often
observed for the ``wrong'' coupling sign (i.e. different from that of
$\eta^\text{True}$). 
The limits on the CI coupling, calculated using the procedure described above, 
are therefore very weak ($\eta^- = -4.4\tev^{-2}$ and $\eta^+ = 4.5\tev^{-2}$). 
To solve this problem, limits for the VA model were calculated by
restricting the fit range to negative or positive couplings, 
corresponding to the lower and upper coupling limit, respectively. 

\begin{figure}[tbp]
\begin{center}
\includegraphics[width=0.9\textwidth]{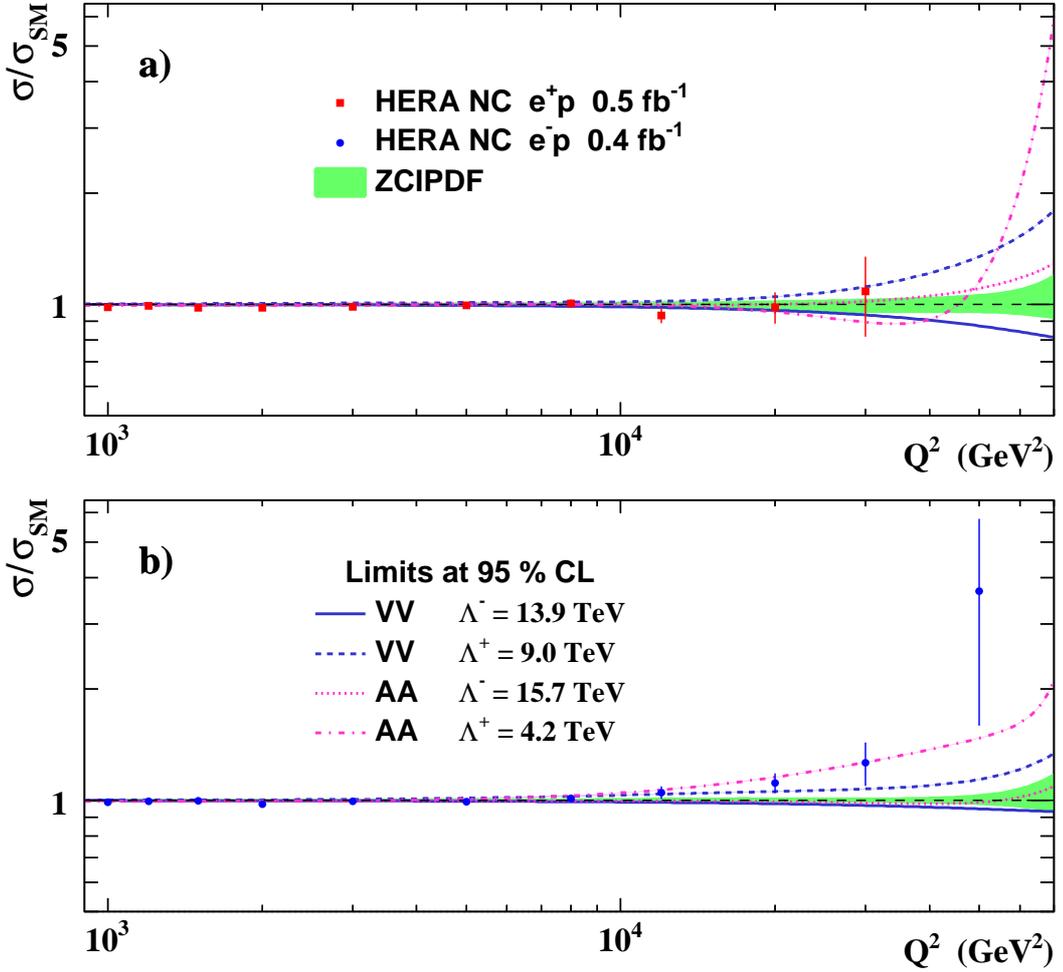}
\end{center}
  \caption{
HERA (a) $e^+p$ and (b) $e^-p$ NC DIS data, relative to the SM
expectations based on the ZCIPDF fit to the HERA inclusive data,
compared to expectations from the  VV and AA contact-interaction models
with the effective mass scale for positive ($\Lambda^{+}$) and negative
($\Lambda^{-}$)  couplings corresponding to the \CL{95} limits.
The same four models are shown on both plots. 
The bands represent the total uncertainty on the ZCIPDF fit predictions. 
         }
  \label{fig-vvaa}
\end{figure}

Compositeness-scale limits calculated taking  modelling uncertainties
into account range from 3.1\,TeV for the X6 model ($\Lambda^{-}$) up
to 17.9\,TeV for the X3 model ($\Lambda^{-}$). 
For the three models mentioned above (AA, X1 and X6), 
when only experimental uncertainties are considered, one sign of the CI
coupling is excluded at \CL{95} and the limits for the coupling and
compositeness scale $\Lambda$ are presented only  
for the other sign.
The effect also persists when modelling uncertainties are 
taken into account for the X1 and X6 scenarios.
In  \fig{vvaa}, the measured $Q^{2}$ spectra of the HERA $e^{+}p$ and $e^{-}p$
data, relative to the SM predictions calculated using ZCIPDF,
are compared with the expectations for the VV and AA
contact-interaction models (as examples) which correspond to the
compositeness limits described above.

\begin{figure}[tbp]
\begin{center}
  \includegraphics[width=0.8\textwidth]{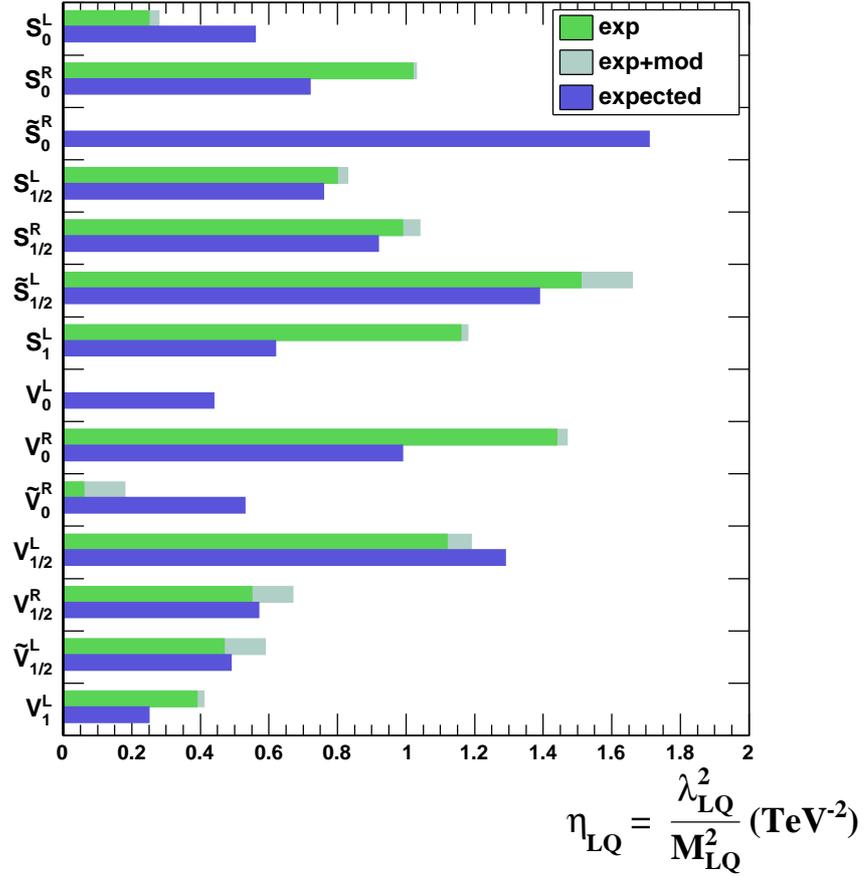}
\end{center}
  \caption{
    Upper \CL{95} limits on the LQ coupling strength, 
    $\eta_\text{LQ} =\lambda^2_\text{LQ}/M^2_\text{LQ}$. 
    Compared are the limits
    calculated without (dark upper bars) and with (light upper bars)  
    modelling uncertainties, and the expected limits (lower bars).
    }
  \label{fig-bars2}
\end{figure}

The LQ coupling values determined from the fit to the HERA inclusive
data, $\eta_\textrm{LQ}^\textrm{Data}$, and the probabilities
$p_\text{SM}$ are summarised in \tab{lq} together with the
coefficients $a^{eq}_{ij}$ describing the CI coupling structure of the
considered LQ models. 
Also shown are the \CL{95} upper limits on the ratio of the Yukawa
coupling to the leptoquark mass, $\lambda_\text{LQ}/M_\text{LQ}$.
Limits calculated without (exp) and with (exp+mod) model and
parameterisation variations are compared with the expected  
\CL{95} limits on $\lambda_\text{LQ}/M_\text{LQ}$  in \fig{bars2}.

For the $S_{1}^{L}$ model, an improvement in the description of the
HERA data can be obtained and the probability of reproducing the fit
result with SM replicas, $p_\text{SM}$,  is below 0.01\%.
For the $V^{R}_{0}$ model, the probability $p_\text{SM}$ is 1.8\%, which means 
that for both models $\eta_\textrm{LQ} = 0$ is excluded at \CL{95}.
When modelling uncertainties are taken into account, the corresponding
$p_\text{SM}$ values increase, but are still below 5\% for both models. 
A probability of less than 5\% is also obtained for the
$\tilde{S}_{0}^{R}$ and $V^{L}_{0}$ models.\footnote{Note 
  that the $\tilde{S}_{0}^{R}$ is related to the $V_{0}^R$ model,
  corresponding to the same CI coupling structure, but with opposite
  sign. The fit results are therefore not independent.}
However, the fit gives unphysical (negative) coupling values so that
both models are excluded at \CL{95}. 

\begin{figure}[tbp]
\begin{center}
\includegraphics[width=0.9\textwidth]{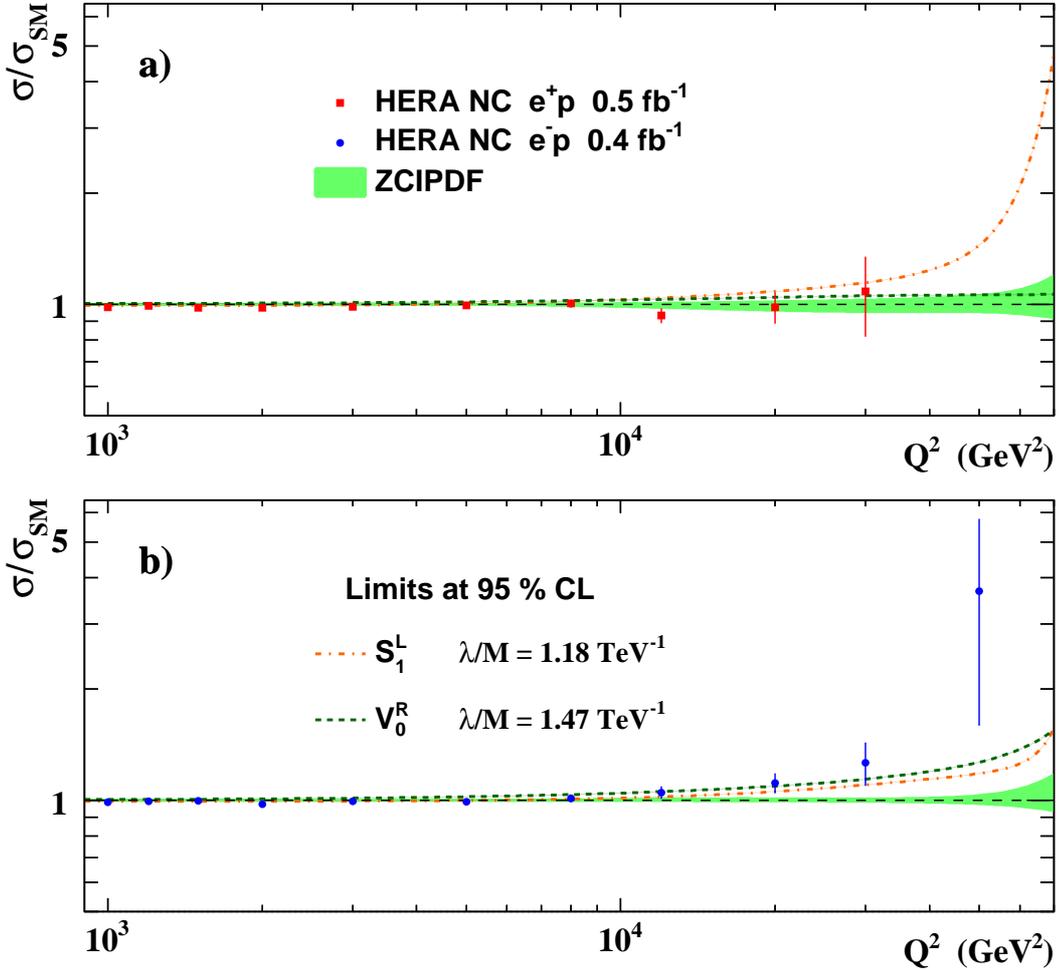}
\end{center}
  \caption{ 
HERA (a) $e^+p$ and (b) $e^-p$ NC DIS data,
relative to the SM expectations based on the ZCIPDF fit to the HERA
inclusive data, compared to expectations from the  $S_{1}^{L}$ and
$V^{R}_{0}$ leptoquark  models with the ratios of the LQ Yukawa couplings
to the LQ mass, $\lambda/M$, corresponding to the \CL{95} limits.
The same two models are shown on both plots. 
The bands represent the total uncertainty  on the ZCIPDF fit predictions. 
 }
  \label{fig-lqsv}
\end{figure}

Assuming the Yukawa coupling value, $\lambda_\text{LQ}=1$, the
corresponding lower limits on the leptoquark mass vary between
0.66$\tev$ for the $\tilde{S}^L_{1/2}$ model and 16$\tev$ for the
$\tilde{V}^R_{0}$ model. 
When modelling uncertainties are included, the limits vary between
0.60$\tev$ and 5.6$\tev$.
In  \fig{lqsv}, the measured $Q^{2}$ spectra of the HERA $e^{+}p$ and $e^{-}p$
data, relative to the SM predictions calculated using ZCIPDF,
are compared with the expectations for the $S_{1}^{L}$ and $V_{0}^R$
leptoquark models which correspond to the limits on the ratio of the
leptoquark Yukawa coupling to the leptoquark mass shown in \fig{bars2}.

Two types of limits can be set on the considered BSM scenarios at
  the LHC.
  Direct searches for LQ pair-production result in $M_\text{LQ}$ limits for
  first-generation leptoquarks in the TeV range
  \cite{Aaboud:2019jcc,Sirunyan:2018btu}.
  These limits do not depend on the leptoquark Yukawa coupling and can
  not be directly compared to the presented HERA results.
  Independent limits on the ratio of the Yukawa coupling to the
  LQ mass, $\lambda_\text{LQ}/M_\text{LQ}$, as well as limits on the
  CI mass scales can be set from the analysis of the dilepton
  production in the Drell--Yan process. 
  The diagram for this process corresponds to that describing NC
  DIS at HERA and the BSM contributions can be searched for in exactly
  the same CI framework.

A comparison of the present results with limits obtained by 
the ATLAS \cite{Aaboud:2017buh} and CMS \cite{Sirunyan:2018ipj}
collaborations at the LHC, based on dilepton data collected at 13 TeV,
is presented in \tab{lhc}. 
Only the four CI models shown were considered in the analyses of the
LHC data. 
It is clear that, for these models, the statistical sensitivity of the
LHC experiments is much higher than that of the HERA inclusive data. 
However, the systematic uncertainties resulting from 
the proton PDFs can be underestimated, as the possible
bias in the parameterisation was not taken into account.

\section{Conclusions}
\label{sec-conclusions}

The HERA combined measurement of inclusive deep inelastic cross
sections in neutral and charged current $e^{\pm} p$ scattering has
been used to search for possible deviations from the Standard Model
predictions within the $eeqq$ contact-interaction approximation.
The procedure was based on a simultaneous fit of PDF 
parameters and the CI coupling.
Limits on the CI couplings and probabilities
for the SM predictions were obtained with Monte Carlo replicas.
These were used to set limits on the CI compositeness scales,
$\Lambda$, and limits on the ratios of the leptoquark Yukawa coupling
to the leptoquark mass, $\lambda_\text{LQ}/M_\text{LQ}$. 

The addition of terms effective at high $Q^2$, such as those found in the
CI and LQ models described above, reduces the tension between high-$Q^2$
and low-$Q^2$ data previously observed in the HERAPDF2.0 analysis.
For the AA, VA, X1 and X6 models, the QCD+CI fits provide improved
descriptions of the HERA inclusive data, corresponding to a difference
from the SM predictions at the level of up to $2.7\,\sigma$ (SM
probability of 0.3\% for the X6 model).
A similar effect is observed for the $S^L_1$ and $V^R_0$ leptoquark
models, which give an improved description of the HERA inclusive data,
corresponding to a difference from the SM predictions at a level of
about $4\,\sigma$ and about $2\,\sigma$, respectively.
These deviations are unlikely to result from statistical
fluctuations alone, but might be explicable by a combination of
modelling uncertainties in the fitting procedure and statistical
fluctuations.
Since an unambiguous deviation from the SM cannot be
established with the HERA data, limits for CI compositeness scales and
LQ mass scales were set that are in the TeV range.

% ----------------------------------------------------------------------------
%       Acknowledgements
% ----------------------------------------------------------------------------

\section*{Acknowledgements}
\label{sec-acknowledgements}

\Zacknowledge

\clearpage

%------------------------------------------------------------------------------
%       Bibliography
%------------------------------------------------------------------------------
{
\bibliographystyle{./inputs/l4z_default3_nodoi}
{\raggedright
\bibliography{./ci_paper.bib,%
              ./inputs/l4z_zeus.bib,%
              ./inputs/l4z_articles.bib}}%
}
\vfill\eject

%------------------------------------------------------------------------------
%       Tables
%------------------------------------------------------------------------------
%------------------------------------------------------------------------------
%       Fit to general CI
%------------------------------------------------------------------------------

\begin{table}[tbp]
  \begin{center}
  \def\arraystretch{1.5}  % For more separation between rows
  \begin{tabular}{|c l@{,}r@{,}r@{, }r|r|c|c|c|c|}
  \multicolumn{10}{c}{\textbf{\large ZEUS}} \\
  \multicolumn{10}{c}{{ HERA $e^\pm p$ 1994--2007 data }}                            \\
\hline
\multicolumn{5}{|c|}{Coupling structure}  &
\multicolumn{4}{c|}{Coupling fit results (TeV$^{-2}$)}  &
\multirow{ 2}{*}{~~~$\Delta\chi^2$~~~}      \\ \cline{6-9}
Model   &~[$\epsilon_{_{LL}}$ & $\epsilon_{_{LR}}$  & $\epsilon_{_{RL}}$ &  $\epsilon_{_{RR}}$]~~   &
$\eta^\textrm{Data}$   & 
$\delta_\textrm{exp}$  & $\delta_\textrm{mod}$  & $\delta_\textrm{tot}$  &             \\
\hline
LL &~[+1  &   0  &   0  &   0]~~  &  ~0.305 & 0.206 & ${}^{+ 0.017}_{- 0.037}$ & ${}^{+ 0.207}_{- 0.209}$ &$-$2.06  \\
RR &~[~~0 &  0   &  0   &  +1]~~  &  ~0.338 & 0.210 & ${}^{+ 0.019}_{- 0.038}$ & ${}^{+ 0.210}_{- 0.213}$ &$-$2.30  \\
LR &~[~~0 & +1   &  0   &   0]~~  & $-$0.084 & 0.247 & ${}^{+ 0.212}_{- 0.060}$ & ${}^{+ 0.325}_{- 0.254}$ &$-$0.12  \\
RL &~[~~0 &  0   & +1   &   0]~~  & $-$0.040 & 0.241 & ${}^{+ 0.198}_{- 0.057}$ & ${}^{+ 0.312}_{- 0.248}$ &$-$0.03  \\
\hline
VV &~[+1  &  +1  &  +1  &  +1]~~  &  ~0.041 & 0.061 & ${}^{+ 0.024}_{- 0.009}$ & ${}^{+ 0.066}_{- 0.062}$ &$-$0.45 \\
AA &~[+1  & $-1$ & $-1$ &  +1]~~  &  ~0.326 & 0.161 & ${}^{+ 0.250}_{- 0.175}$ & ${}^{+ 0.297}_{- 0.238}$ &$-$4.67 \\[+1mm]
\multirow{ 2}{*}{VA} &
\multicolumn{1}{l@{}}{\multirow{ 2}{*}{~[+1,}} &
\multicolumn{1}{r@{}}{\multirow{ 2}{*}{$-1$,}} &
\multicolumn{1}{r@{}}{\multirow{ 2}{*}{+1, }} &
\multirow{ 2}{*}{$-1$]~~}
                                  & $-$0.594 & 0.225 & ${}^{+ 0.028}_{- 0.120}$ & ${}^{+ 0.227}_{- 0.255}$ &$-$1.21  \\[-1mm]
\multicolumn{5}{|c|}{}
                                  &  ~0.676 & 0.200 & ${}^{+ 0.078}_{- 0.019}$ & ${}^{+ 0.215}_{- 0.201}$ &$-$3.25  \\[+1mm]
X1 &~[+1  & $-1$ &   0  &   0]~~  &  ~0.682 & 0.267 & ${}^{+ 0.339}_{- 0.243}$ & ${}^{+ 0.432}_{- 0.361}$ &$-$5.52  \\
X2 &~[+1  &   0  &  +1  &   0]~~  &  ~0.089 & 0.121 & ${}^{+ 0.046}_{- 0.017}$ & ${}^{+ 0.129}_{- 0.122}$ &$-$0.52  \\
X3 &~[+1  &   0  &  0  &   +1]~~  &  ~0.158 & 0.108 & ${}^{+ 0.009}_{- 0.019}$ & ${}^{+ 0.109}_{- 0.110}$ &$-$2.09  \\
X4 &~[~~0  &  +1  &  +1  &   0]~~~& $-$0.029 & 0.116 & ${}^{+ 0.098}_{- 0.026}$ & ${}^{+ 0.151}_{- 0.119}$ &$-$0.06  \\
X5 &~[~~0  &  +1  &  0   &  +1]~~~&  ~0.079 & 0.123 & ${}^{+ 0.052}_{- 0.018}$ & ${}^{+ 0.133}_{- 0.124}$ &$-$0.41  \\
X6 &~[~~0  &  0  &  +1  &  $-1$]~~~&$-$0.786 & 0.274 & ${}^{+ 0.192}_{- 0.295}$ & ${}^{+ 0.334}_{- 0.402}$ &$-$6.01  \\
\hline
\end{tabular}
  \end{center}
  \caption{
     Relations between couplings 
  $[\epsilon_{LL},\epsilon_{LR}, \epsilon_{RL}, \epsilon_{RR}]$ for the considered
     compositeness models and the CI coupling values, $\eta^\textrm{Data}$,  
     determined from the simultaneous QCD+CI fit to the HERA inclusive data; $\delta_\textrm{exp}$, 
     $\delta_\textrm{mod}$ and $\delta_\textrm{tot}$, represent the experimental, modelling and total uncertainties, respectively. 
     Also shown is the change of the $\chi^2$ value relative to the fit performed 
     without the CI contribution, $\Delta\chi^2=\chi^2_\textrm{SM+CI}-\chi^2_\textrm{SM}$.
     For the VA model, two minima in the $\chi^2$ distribution were considered,
     separately for negative and for positive coupling values (see text for details). 
        }
  \label{tab-cifit}
\end{table}

%------------------------------------------------------------------------------
%     Fit to leptoquarks
%------------------------------------------------------------------------------
\begin{table}[btp]
  \begin{center}
  \def\arraystretch{1.3}  % For more separation between rows
    \begin{tabular}{|cl|r|c|c|c|c|}
      \multicolumn{7}{c}{\textbf{\large ZEUS}} \\
      % \hline
      \multicolumn{7}{c}{{HERA $e^\pm p$ 1994--2007 data}} \\

    \hline

    \multirow{2}{*}{Model} & \multirow{2}{*}{Coupling Structure}
             & \multicolumn{4}{c|}{Coupling fit results (TeV$^{-2}$)}  
             & \multirow{ 2}{*}{~~~$\Delta\chi^2$~~~}     \\ \cline{3-6}
      &  &  $\eta_\textrm{LQ}^\textrm{Data}$   
      & $\delta_\textrm{exp}$  & $\delta_\textrm{mod}$  & $\delta_\textrm{tot}$  &  \\
      \hline
      $S_{0}^L$
      &  $a^{eu}_{_{LL}}=+\frac{1}{2}$
      &$-$0.258 & 0.196 & ${}^{+ 0.034}_{- 0.036}$ & ${}^{+ 0.199}_{- 0.199}$
      &$-$1.56
      \\
      $S_{0}^R$
      &  $a^{eu}_{_{RR}}=+\frac{1}{2}$
      & ~0.533 & 0.331 & ${}^{+ 0.034}_{- 0.061}$ & ${}^{+ 0.332}_{- 0.336}$
      &$-$2.53
      \\
      $\tilde{S}_{0}^{R}$
      &  $a^{ed}_{_{RR}}=+\frac{1}{2}$
      &$-$2.561 & 1.115 & ${}^{+ 0.323}_{- 0.221}$ & ${}^{+ 1.161}_{- 1.137}$
      &$-$3.98
      \\
      $S_{1/2}^L$
      &  $a^{eu}_{_{LR}}=-\frac{1}{2}$
      & ~0.054 & 0.341 & ${}^{+ 0.075}_{- 0.280}$ & ${}^{+ 0.349}_{- 0.441}$
      &$-$0.02
      \\
      $S_{1/2}^R$
      &  $a^{ed}_{_{RL}}=a^{eu}_{_{RL}}=-\frac{1}{2}$
      & ~0.112 & 0.491 & ${}^{+ 0.118}_{- 0.412}$ & ${}^{+ 0.505}_{- 0.641}$
      &$-$0.05
      \\
      $\tilde{S}_{1/2}^{L}$
      &  $a^{ed}_{_{LR}}=-\frac{1}{2}$
      & ~0.464 & 1.371 & ${}^{+ 0.925}_{- 0.264}$ & ${}^{+ 1.654}_{- 1.396}$
      &$-$0.10
      \\
      $S_{1}^{L}$
      &  $a^{ed}_{_{LL}}=+1, \; a^{eu}_{_{LL}}=+\frac{1}{2}$
      & ~0.974 & 0.203 & ${}^{+ 0.043}_{- 0.337}$ & ${}^{+ 0.207}_{- 0.393}$
      &$-$11.10
      \\
      \hline
      $V_{0}^L$
      &  $a^{ed}_{_{LL}}=-1$
      &$-$0.325 & 0.116 & ${}^{+ 0.030}_{- 0.101}$ & ${}^{+ 0.120}_{- 0.154}$
      &$-$6.17
      \\
      $V_{0}^R$
      &  $a^{ed}_{_{RR}}=-1$
      & ~1.280 & 0.558 & ${}^{+ 0.111}_{- 0.163}$ & ${}^{+ 0.568}_{- 0.581}$
      &$-$3.98
      \\
      $\tilde{V}_{0}^{R}$
      &  $a^{eu}_{_{RR}}=-1$
      &$-$0.267 & 0.165 & ${}^{+ 0.030}_{- 0.017}$ & ${}^{+ 0.168}_{- 0.166}$
      &$-$2.53
      \\
      $V_{1/2}^L$
      &  $a^{ed}_{_{LR}}=+1$
      &$-$0.232 & 0.685 & ${}^{+ 0.132}_{- 0.460}$ & ${}^{+ 0.698}_{- 0.825}$
      &$-$0.10
      \\
      $V_{1/2}^R$
      &  $a^{ed}_{_{RL}}=a^{eu}_{_{RL}}=+1$
      &$-$0.056 & 0.246 & ${}^{+ 0.206}_{- 0.059}$ & ${}^{+ 0.320}_{- 0.253}$
      &$-$0.05
      \\
      $\tilde{V}_{1/2}^{L}$
      &  $a^{eu}_{_{LR}}=+1$
      &$-$0.027 & 0.171 & ${}^{+ 0.139}_{- 0.038}$ & ${}^{+ 0.220}_{- 0.175}$
      &$-$0.02
      \\
      $V_{1}^{L}$
      &  $a^{ed}_{_{LL}}=-1, \; a^{eu}_{_{LL}}=-2 $
      & ~0.029 & 0.077 & ${}^{+ 0.015}_{- 0.013}$ & ${}^{+ 0.079}_{- 0.079}$
      &$-$0.14
      \\
      \hline
    \end{tabular}
  \end{center}
  \caption{Coefficients $a^{eq}_{ij}$ defining the effective
    LQ couplings in the CI limit,
    $M_{LQ}\gg \sqrt{s}$, and the coupling values, $\eta_\textrm{LQ}^\textrm{Data}$, 
    determined from the simultaneous QCD+CI fit to the HERA inclusive data, for different 
    models of scalar (upper part of the table) and vector (lower part) LQs;  
    $\delta_\textrm{exp}$, $\delta_\textrm{mod}$ and $\delta_\textrm{tot}$, 
    represent the experimental, modelling and total uncertainties, respectively. 
     Also shown is the change of the $\chi^2$ value relative to the fit performed 
     without the LQ contribution, $\Delta\chi^2=\chi^2_\textrm{SM+LQ}-\chi^2_\textrm{SM}$. 
  }
  \label{tab-lqfit}
\end{table}

%------------------------------------------------------------------------------
%       Uncertainties
%------------------------------------------------------------------------------

\begin{table}[tbp]
\renewcommand*{\arraystretch}{1.2}
\addtolength{\tabcolsep}{8mm}
\centerline{
\begin{tabular}{|@{ }l|l|l|l|}
%\vspace{-1.0cm}
\hline
\multicolumn{1}{|c|}{Variation} &
\multicolumn{1}{@{ }c@{ }|}{Nominal Value} &
\multicolumn{1}{@{ }c@{ }|}{Lower Limit} &
\multicolumn{1}{@{ }c@{ }|}{Upper Limit}  \\
\hline
$Q^2_\text{min}$ [GeV$^2$] & 3.5  & 2.5   & 5.0   \\
\hline
charm mass parameter $M_c$ [GeV]     & 1.47    & 1.41        & 1.53   \\
beauty mass parameter $M_b$ [GeV]     & 4.5    & 4.25 & 4.75  \\
\hline
sea strange fraction $f_s$           & 0.4      & 0.3 & 0.5   \\
\hline
starting scale $\mu^2_{f_{0}}$ [GeV$^2$]       & 1.9      & 1.6 & 2.2  \\
\hline
\end{tabular}}
\addtolength{\tabcolsep}{-8mm}
\caption{Input parameters for the fit and the variations 
considered to evaluate model and parameterisation
uncertainties.  
}
\label{tab-model}
\end{table}

%------------------------------------------------------------------------------
%       general CI
%------------------------------------------------------------------------------

\begin{landscape}

\begin{table}[tbp]
  \begin{center}
 \def\arraystretch{1.08}  % For more separation between rows
  \begin{tabular}{|c|r|c|cc|cc|cc|cc|cc|}
  \multicolumn{13}{c}{\textbf{\large ZEUS}} \\
  \multicolumn{13}{c}{{ HERA $e^\pm p$ 1994--2007 data }}                            \\
\hline
 &  &  &
\multicolumn{6}{c|}{95\% C.L. coupling limits (TeV$^{-2}$)}  & 
\multicolumn{4}{c|}{95\% C.L. mass scale limits (TeV)}                \\   \cline{4-13}
Model
&
$\eta^\textrm{Data}$~~~    &
$p_\textrm{SM}$                &
\multicolumn{2}{c|}{Measured (exp)}     &
\multicolumn{2}{c|}{Measured (exp+mod)}     &
\multicolumn{2}{c|}{Expected}     &
\multicolumn{2}{c|}{Measured (exp+mod)}     &
\multicolumn{2}{c|}{Expected}     \\
   &
(TeV$^{-2}$)    &
(\%)              &
~~~$\eta^-$  &  $\eta^+$  &
~~~$\eta^-$  &  $\eta^+$  &
~~~$\eta^-$  &  $\eta^+$  &
~~~$\Lambda^{-}$  &  $\Lambda^{+}$  &
~~~$\Lambda^{-}$  &  $\Lambda^{+}$   \\
\hline
LL &  ~0.305 & 7.0 &$-$0.033 & ~0.610 &$-$0.077 & ~0.616  & $-$0.367 & 0.319  &  12.8 & 4.5  & 5.9  &  6.3  \\
RR &  ~0.338 & 5.9 &$-$0.017 & ~0.649 &$-$0.058 & ~0.656  & $-$0.390 & 0.337  &   14.7 & 4.4  & 5.7  &  6.1  \\
LR & $-$0.084 & 34  &$-$0.514 & ~0.250 &$-$0.565 & ~0.413  & $-$0.388 & 0.313  &  4.7 & 5.5    & 5.7  &  6.3   \\
RL & $-$0.040 & 42  &$-$0.464 & ~0.299 &$-$0.503 & ~0.444  & $-$0.397 & 0.302  &  5.0 & 5.3    & 5.6  &  6.5  \\
\hline
VV &  ~0.041 & 25  &$-$0.058 & ~0.135 &$-$0.065 & ~0.155  &$-$0.101 & 0.097 &   13.9 & 9.0   & 11.2 &  11.4   \\
AA &  ~0.326 & 0.6 &         & ~0.530 &$-$0.051 & ~0.700  &$-$0.200 & 0.207 &    15.7 & 4.2   & 7.9  & 7.8   \\[+1mm]
\multirow{ 2}{*}{VA}              & $-$0.594   & 5.8  
                                  & \multirow{2}{*}{$-$0.888} & \multirow{2}{*}{0.947}    
                                  &  \multirow{2}{*}{$-$0.969} & \multirow{2}{*}{0.997}
                                  & \multirow{2}{*}{$-$0.723} &\multirow{ 2}{*}{0.719} 
                                  & \multirow{2}{*}{ 3.6   } & \multirow{2}{*}{3.5} 
                                  & \multirow{2}{*}{4.2}  & \multirow{2}{*}{4.2}  \\[-1mm]
                                  &  ~0.676 & 2.5 &  &  &   &  &  &  
                                  &   &  &  &  \\[+1mm]
X1 &  ~0.682 & 0.4 &         & ~1.020 &        & ~1.230  & $-$0.435 & 0.418  &      & 3.2  &  5.4  &  5.5 \\
X2 &  ~0.089 & 24  &$-$0.113 & ~0.269 &$-$0.125 & ~0.310  & $-$0.206 & 0.184  &  10.4 & 6.4  &  7.8 &  8.3   \\
X3 &  ~0.158 & 7.3 &$-$0.018 & ~0.320 &$-$0.039 & ~0.324  & $-$0.183 & 0.166  &  17.9 & 6.2  &  8.3 &  8.7   \\
X4 & $-$0.029 & 39  &$-$0.230 & ~0.144 &$-$0.243 & ~0.223  & $-$0.194 & 0.170  &   7.2 & 7.5  &  8.0 &  8.6  \\
X5 &  ~0.079 & 27  &$-$0.129 & ~0.263 &$-$0.138 & ~0.303  & $-$0.212 & 0.188  &   9.5 & 6.4  &  7.7 &  7.7   \\
X6 & $-$0.786 & 0.3 &$-$1.130 &        &$-$1.310 &         & $-$0.454 & 0.415  &   3.1 &  &   5.3 &   5.5  \\
\hline
\end{tabular}
  \end{center}
  \caption{
     Contact-interaction coupling values determined from the
     fit to the HERA inclusive data, $\eta^\textrm{Data}$, and probabilities to obtain 
     larger absolute coupling values from the fit to the SM replica, $p_\textrm{SM}$,
     for the considered CI models.
     Also shown are the \CL{95} limits on the CI couplings obtained 
     from the presented analysis without (exp) and with (exp+mod)
     model and parameterisation variations. 
     Lower and upper coupling limits, $\eta^-$ and $\eta^+$, are calculated separately 
     for negative and positive coupling values, respectively.
     The \CL{95} upper  
     limits on the compositeness scale, $\Lambda^{+}$ and $\Lambda^{-}$, correspond to 
     the scenarios with positive  and negative coupling values, respectively.
     The same coupling structure applies to all quarks. Only positive coupling
     values are allowed at \CL{95} for the X1 model, and for the AA model when
     modelling uncertainties are not taken into account,
     while for the X6 model only negative coupling values are allowed.
     For the VA model, the fit range is restricted to negative or positive couplings
    for lower and upper limit calculations, respectively (see text for details).     }
  \label{tab-ci}
\end{table}

\end{landscape}

%------------------------------------------------------------------------------
%       Lepto quarks
%------------------------------------------------------------------------------
\addtolength{\tabcolsep}{-1mm}

\begin{table}[btp]
  \begin{center}
  \def\arraystretch{1.5}  % For more separation between rows
    \begin{tabular}{|cl|r|c|c|c|c|}
      \multicolumn{7}{c}{\textbf{\large ZEUS}} \\
      % \hline
      \multicolumn{7}{c}{{HERA $e^\pm p$ 1994--2007 data}} \\

    \hline

      &  &  & & \multicolumn{3}{c|}{$\lambda_\textrm{LQ}/M_\textrm{LQ}$  95\%~C.L. limits (TeV$^{-1}$)}   \\
      \cline{5-7}
      Model & Coupling Structure & $\eta^\textrm{Data}_\textrm{LQ}$~~ & ~~~$p_\textrm{SM}$~~~ &
        \multicolumn{2}{c|}{Measured} & \multirow{2}{*}{Expected} \\

      \cline{5-6}
      &  &  (TeV$^{-2}$)  & (\%)  &  ~~~~(exp)~~~~ & (exp+mod) &  \\
      \hline
      $S_{0}^L$
      &  $a^{eu}_{_{LL}}=+\frac{1}{2}$
      &$-$0.258
      & 9.0 %   91.6
      & 0.25 & 0.28
      & 0.56
      \\
      $S_{0}^R$
      &  $a^{eu}_{_{RR}}=+\frac{1}{2}$
      & ~0.533
      & 5.5
      & 1.02 & 1.03
      & 0.72
      \\
      $\tilde{S}_{0}^{R}$
      &  $a^{ed}_{_{RR}}=+\frac{1}{2}$
      &$-$2.561
      & 1.8  %  Changed to V_o_R probability  (was 98.5)
      &    &     %  Better not to quote 0...
      & 1.71
      \\
      $S_{1/2}^L$
      &  $a^{eu}_{_{LR}}=-\frac{1}{2}$
      & ~0.054
      & 43
      & 0.80  &  0.83
      & 0.76
      \\
      $S_{1/2}^R$
      &  $a^{ed}_{_{RL}}=a^{eu}_{_{RL}}=-\frac{1}{2}$
      & ~0.112
      & 39
      & 0.99 & 1.04
      & 0.92
      \\
      $\tilde{S}_{1/2}^{L}$
      &  $a^{ed}_{_{LR}}=-\frac{1}{2}$
      & ~0.464
      & 38
      & 1.51 & 1.66
      & 1.39
      \\
      $S_{1}^{L}$
      &  $a^{ed}_{_{LL}}=+1, \; a^{eu}_{_{LL}}=+\frac{1}{2}$
      & ~0.974
      & $< 0.01$
      & 1.16 & 1.18
      & 0.62
      \\
      \hline
      $V_{0}^L$
      &  $a^{ed}_{_{LL}}=-1$
      &$-$0.325
      & 0.5
      &    &     %  Better not to quote 0...
      & 0.44
      \\
      $V_{0}^R$
      &  $a^{ed}_{_{RR}}=-1$
      & ~1.280
      & 1.8  % Changed for consistency with the plot 2.1
      & 1.44 & 1.47
      & 0.99
      \\
      $\tilde{V}_{0}^{R}$
      &  $a^{eu}_{_{RR}}=-1$
      &$-$0.267
      & 5.5
      & 0.06 & 0.18
      & 0.53
      \\
      $V_{1/2}^L$
      &  $a^{ed}_{_{LR}}=+1$
      &$-$0.232
      & 38
      & 1.12 & 1.19
      & 1.29
      \\
      $V_{1/2}^R$
      &  $a^{ed}_{_{RL}}=a^{eu}_{_{RL}}=+1$
      &$-$0.056
      & 39
      & 0.55 & 0.67
      & 0.57
      \\
      $\tilde{V}_{1/2}^{L}$
      &  $a^{eu}_{_{LR}}=+1$
      &$-$0.027
      & 43
      & 0.47 & 0.59
      & 0.49
      \\
      $V_{1}^{L}$
      &  $a^{ed}_{_{LL}}=-1, \; a^{eu}_{_{LL}}=-2 $
      & ~0.029
      & 32
      & 0.39 & 0.41
      & 0.25
      \\
      \hline
    \end{tabular}
  \end{center}
  \caption{Coefficients $a^{eq}_{ij}$ defining the effective
    leptoquark couplings in the contact-interaction limit,
    $M_{LQ}\gg \sqrt{s}$, coupling values determined from the
     fit to the HERA inclusive data, $\eta_\textrm{LQ}^\textrm{Data}$, and the upper limits on
    the Yukawa coupling to the leptoquark mass ratio, $\lambda_{LQ}/M_{LQ}$,
    for different models of scalar (upper part of the table)
    and vector (lower part) leptoquarks.
    Limits calculated without (exp) and with (exp+mod)
    modelling uncertainties are compared with the 
    expected limits.
    For the $\tilde{S}_{0}^{R}$ and $V_{0}^{L}$ models,
    all positive coupling values are excluded at \CL{95}.
  }
  \label{tab-lq}
\end{table}

%-------------------------------------------------------------------
%  Comparison with LHC
%-------------------------------------------------------------------

\addtolength{\tabcolsep}{-1mm}

\begin{table}[tbp]
  \begin{center}
  \def\arraystretch{1.5}  % For more separation between rows
   \begin{tabular}{|c@{~~~[}r@{,}r@{,}r@{,}r|cc|cc|cc|}
  \multicolumn{11}{c}{\textbf{\large ZEUS}} \\
\hline
  \multicolumn{5}{|c|}{}  &  
  \multicolumn{6}{c|}{\CL{95} limits (\tev)}  \\
\cline{6-11}
  \multicolumn{5}{|c|}{Coupling structure}  &  
  \multicolumn{2}{c|}{HERA}     &
  \multicolumn{2}{c|}{ATLAS}    &
  \multicolumn{2}{c|}{CMS}   \\
~Model  & 
 $\epsilon_{_{LL}}$ & $\epsilon_{_{LR}}$  &
                 $\epsilon_{_{RL}}$ &  $\epsilon_{_{RR}}$]~~ &
~~~~$\Lambda^{-}$~~~~  &  ~~~~$\Lambda^{+}$~~~~ &
~~~~$\Lambda^{-}$~~~~  &  ~~~~$\Lambda^{+}$~~~~ &
~~~~$\Lambda^{-}$~~~~  &  ~~~~$\Lambda^{+}$~~~~ \\
\hline
LL 
&  +1  &  0  &  0  &  0 ]~~
 & 12.8  & 4.5
 &  24 & 37
 & 16.8  & 24.0   \\
RR 
 &  0  &  0  &  0  &  +1 ]~~      
 & 14.7  &  4.4
 & 26  &  33 
 & 16.9 & 23.8  \\
LR
&  0  &  +1  &  0  &  0 ]~~
 & 4.7  & 5.5
 & 26 &  33 
 & 21.3  & 26.4   \\
RL 
&  0  &  0  &  +1  &  0 ]~~
 & 5.0  & 5.3
 & 26  & 33  
 &   &    \\
\hline
    \end{tabular}
  \end{center}
  \caption{Comparison of the \CL{95} limits on 
         the compositeness scale, $\Lambda$,
         obtained from the ZEUS analysis of the HERA inclusive data
         with limits on $eeqq$ CI resulting from the analysis of the
         dilepton mass spectra at 13\,TeV LHC presented
         by the ATLAS Collaboration \protect\cite{Aaboud:2017buh} and
         by the CMS Collaboration \protect\cite{Sirunyan:2018ipj}.
         For the ATLAS experiment, limits that were obtained
         in a Bayesian framework with an assumed uniform positive
         prior in $1/\Lambda^{2}$ are shown.
         }
  \label{tab-lhc}
\end{table}

\addtolength{\tabcolsep}{2mm}

%
%------------------------------------------------------------------------------
%       Figures
%------------------------------------------------------------------------------
%-------------------------------------------------------------------------------
%       CI fit vs data
%-------------------------------------------------------------------------------

%-------------------------------------------------------------------------------
%       LQ fit vs data
%-------------------------------------------------------------------------------

%------------------------------------------------------------------------------
%       Summary plot (limit bars)
%------------------------------------------------------------------------------

%-------------------------------------------------------------------------------
%       CI limits vs data
%-------------------------------------------------------------------------------

%------------------------------------------------------------------------------
%       Summary plot (limit bars)
%------------------------------------------------------------------------------

%-------------------------------------------------------------------------------
%       LQ limits vs data
%-------------------------------------------------------------------------------

%
%       ... that's it
%
\end{document}